\documentclass[iop,apj,12pt,numberedappendix]{emulateapj}
\usepackage{times}
\usepackage[normalem]{ulem}
\usepackage{graphicx}
\usepackage{natbib}
\usepackage{amsfonts}
\usepackage{amsmath}
\usepackage{multirow}
\usepackage{enumerate}
\usepackage{cancel}
\usepackage{slashed}
\usepackage{comment}
\usepackage{float}
\usepackage{verbatim}
\usepackage{wrapfig}
\usepackage[usenames,dvipsnames]{xcolor}
\usepackage[plainpages=false, colorlinks=true, anchorcolor=blue, linkcolor=blue, citecolor=blue, bookmarks=false]{hyperref}
\newcommand{\rqm}{{%
\declareslashed{}{\text{--}}{0.08}{-0.5}{I}\slashed{I}}}

\newcommand{\be}{\begin{eqnarray}}
\newcommand{\ee}{\end{eqnarray}}

\newcommand{\shortauth}{Morozova et al.}
\newcommand{\slugcom}{Accepted for publication in The Astrophysical Journal}
\slugcomment{\slugcom}

\newcommand{\GG}[1]{}

\lefthead{\sc \footnotesize \slugcom \hfill \shortauth}
\righthead{\sc \footnotesize \slugcom \hfill \shortauth}

\begin{document}

\title{The gravitational wave signal from core-collapse supernovae}

\author{Viktoriya Morozova\altaffilmark{1}}
\author{David Radice\altaffilmark{1,2}}
\author{Adam Burrows\altaffilmark{1}}
\author{David Vartanyan\altaffilmark{1}}
\altaffiltext{1}{Department of Astrophysical Sciences,
  Princeton University, Princeton, NJ 08544, USA, 
  vsg@astro.princeton.edu}
\altaffiltext{2}{Schmidt Fellow, Institute for Advanced Study, 
1 Einstein Drive, Princeton, NJ 08540}

\begin{abstract}

We study gravitational waves (GWs) from a set of two-dimensional multi-group
neutrino radiation hydrodynamic simulations of core-collapse supernovae (CCSNe).
Our goal is to systematize the current knowledge about the post-bounce CCSN GW 
signal and recognize the templatable features that could be used by the ground-based
laser interferometers. We demonstrate that starting from $\sim$400$\,{\rm ms}$ 
after core bounce the dominant GW signal represents the fundamental quadrupole ($l=2$)
oscillation mode (f-mode) of the proto-neutron star (PNS), which can be accurately reproduced
by a linear perturbation analysis of the angle-averaged PNS profile. Before that, in the time
interval between $\sim$200 and $\sim$400$\,{\rm ms}$ after bounce, the dominant mode has 
two radial nodes and represents a g-mode. We associate the high-frequency
noise in the GW spectrograms above the main signal with p-modes, while below the dominant
frequency there is a region with very little power.
The collection of models presented here summarizes the dependence of the CCSN GW
signal on the progenitor mass, equation of state, many-body corrections to the 
neutrino opacity, and rotation. Weak dependence of the dominant GW frequency
on the progenitor mass motivates us to provide a simple fit for it as a function of
time, which can be used as a prior when looking for CCSN candidates in the LIGO data.

\end{abstract}

	
  
\section{Introduction}

After decades of development, the ground-based laser 
interferometers LIGO and Virgo have detected the 
gravitational-wave (GW) signal of merging binary systems of black holes
\citep{abbott:16,abbott:17a} and neutron stars \citep{abbott:17b}. 
The latter event was especially interesting, 
because it was subsequently observed in the X-ray, UV, optical, 
infrared, and radio bands \citep{abbott:17c}. 
Detection of the GWs from a galactic 
core-collapse supernova (CCSN),
potentially accompanied by detection of neutrinos and
electromagnetic observations in all available bands, could be the next 
major breakthrough. Given that the estimate for the galactic CCSNe rate is
$3.2^{+7.3}_{-2.6}$ per century \citep{li:11,adams:13}, and the 
youngest known galactic CCSN remnant is $\sim$100 years old
\citep{reynolds:08,borkowski:17}, chances are high that
we will not have long to wait.

Our ability to recognize the CCSN GW signal and extract it from the 
nonstationary and non-Gaussian background noise of the detectors largely depends 
on our knowledge of the signal's time-frequency structure 
\citep[see, for example,][and references therein]{hayama:15,gossan:16}.
Attempts to characterize the GW emission from CCSNe started
in 1960s with the analytical estimates of \citet{wheeler:66} and evolved 
into the fully relativistic
multidimensional numerical simulations of the early 2000s (see reviews of 
\citealt{ott:09} and \citealt{kotake:13}). Indeed, since a spherically-symmetric
star does not emit GWs, and the CCSN mechanism relies on a complex
hydrodynamical evolution of the stellar core, including 
neutrino interactions, fluid instabilities, and
shocks, a two-dimensional (2D) radiation-hydrodynamics code is the 
minumum capability required to model the CCSN GW signal.
Previous studies of the GW signal from 2D CCSN models may be found,
for example, in \citet{marek:09,murphy:09,kotake:09,mueller:13}
and \citet{cedraduran:13}.
Once computationally unaffordable, general-relativistic 3D simulations
of CCSNe with state-of-the-art neutrino physics have been recently 
performed by a number of groups (\citealt{melson:15a,melson:15b,lentz:15,mueller:17,ott:17};
see also \citealt{takiwaki:12,ott:13,takiwaki:14,mueller:15b,roberts:16,pan:17}). 
\citet{yakunin:17,andresen:17,kuroda:16} and \citet{kuroda:17} have provided the
GW signal from their most recent 3D simulations 
\citep[for earlier work, see][]{fryer:04,scheidegger:08,scheidegger:10,mueller:12,ott:12,kuroda:14}.
Here, we show the results of our current 2D study of the GW signals from
the recent \textsc{Fornax} models, building upon the
earlier efforts of our group to simulate CCSN explosions 
\citep{dolence:15,skinner:16,burrows:16,radice:17}.

One of the main difficulties in the study of the GW signal of CCSNe
is related to the stochasticity of the processes responsible for its
generation. For example, the early $\sim$100$\,{\rm Hz}$ signal arising in 
the first tens of milliseconds after core bounce is 
commonly associated with the shock oscillations driven by
prompt convection \citep{marek:09,murphy:09,mueller:13,yakunin:10,yakunin:15}, 
which makes its parameterization very difficult. In the next few hundred 
milliseconds, a
stronger signal follows with the frequency gradually increasing in the range 
$300-1000\,{\rm Hz}$. This signal is usually associated with the surface g-modes
of the newly formed proto-neutron star (PNS) excited by the downflows from
the postshock convection region or from convection inside the PNS itself
\citep{murphy:09,marek:09,mueller:13}.
Some features of the GW signal are known to be associated with the
standing accretion shock instability (SASI) 
\citep{cedraduran:13,kuroda:16,andresen:17,pan:17}.

A number of attempts has been made to systematize
and identify the features of the CCSNe GW signal by means of
asteroseismology, specifically, applying linear perturbation analysis to the PNS and its
surrounding region \citep{fuller:15,sotani:16,torres:17,camelio:17}. For example, 
\citet{fuller:15} showed that the
fundamental quadrupolar oscillation mode of the PNS may be responsible
for the early post-bounce signal of the rapidly rotating core. Recently,
\citet{torres:17} presented a relativistic formalism to identify the eigenfrequencies 
of the PNS and its surrounding postshock region, which they used to analyze the
rotating 2D CCSN model from \citet{cedraduran:13}. In the current study, we go one 
step further
and relax the Cowling approximation used in \citet{torres:17}. After doing so, we can
firmly relate the dominant component of the GW signal from our models with the 
fundamental f-mode quadrupole oscillation of the PNS. This association holds for
different progenitor masses, equations of state (EOS), and numerical prescriptions
for gravity and neutrino interactions used in our study.

We find that the dominant GW frequency depends weakly on the progenitor 
zero-age main-sequence (ZAMS) mass, without any clear systematic trend.
Instead, it is sensitive to the EOS and the details of neutrino microphysics. Motivated
by its simple time evolution, we fit the dominant GW frequency as a function of time 
with a quadratic polynomial, which can be
used as a prior in the GW data analysis, when looking for the CCSN candidates.
We identify a new feature in the form of a power `gap' across the GW spectrogram,
which, if proven physical, may provide some information about the structure of the
inner PNS core. We study the influence of rotation on the GW signal and find that, 
while increasing the power of the core bounce signal, it may weaken the GW emission
in the post-bounce phase.

The paper is organized in the following way. Section~\ref{setup} outlines our numerical setup and
summarizes the CCSN models used in our study. In Section~\ref{example}, we present an
example of the GW spectrogram obtained and describe its key features
common between all our models. In Section~\ref{analysis}, we explain the physical origin of some
of these features by means of the linear perturbation analysis. Section~\ref{dependences} is
devoted to the comparison of the spectrograms from different simulations, which
shows the key dependences of the GW signal on the parameters of the models
and the details of the numerical setup. Discussion and conclusions are given in 
Section~\ref{discussion}. For simplicity, in the sections describing the linear analysis 
(Section~\ref{analysis} and Appendix~\ref{Ap1}), we use the geometrized system
of units $G=c=1$, where $c$ is the speed of light and $G$ is Newton's gravitational
constant. In other sections, $G$ and $c$ are shown explicitly in the equations.

\section{Numerical setup}
\label{setup}

\begin{table*}
\renewcommand{\arraystretch}{1.3}
\centering
\caption{Summary of the models shown in this study. \label{tab:models}}
\begin{tabular}{lccccccccl}\hline \hline
Progenitor & EOS & Inner angular & Gravity & Many-body & Explosion & Simulation & $E_{GW}$ (matter) & $E_{GW}$ (neutrino) & Label \\
mass $[M_{\odot}]$ &  &  velocity $\Omega_0\,[{\rm rad}/{\rm s}]$ & solver  & corrections & status & time [s] & $[10^{-8}\,M_{\odot}c^2]$ & $[10^{-8}\,M_{\odot}c^2]$ &    \\
\hline
$10$ & LS220 & 0 & monopole & yes & no & 1.22 & 0.22 & 0.001 & \texttt{M10\_LS220}   \\ 

 & LS220 & 0 & monopole & no & no & 2.15 & 0.23 & 0.001 & \texttt{M10\_LS220\_no\_manybody}    \\ 

 & SFHo & 0 & monopole  & yes & yes & 1.50 & 1.65 & 0.013 & \texttt{M10\_SFHo}   \\ 

 & DD2 & 0 & monopole & yes & no & 1.66 & 0.16 & 0.001 & \texttt{M10\_DD2}    \\ 

\hline

$13$ & SFHo & 0 & monopole & yes & no & 1.36 & 1.00 & 0.003 & \texttt{M13\_SFHo}    \\ 

 & SFHo & 0 & multipole & yes & no & 0.85 & 0.65 & 0.003 & \texttt{M13\_SFHo\_multipole}    \\ 

 & SFHo & 0.2 & multipole & yes & yes & 1.00 & 0.27 & 0.010 & \texttt{M13\_SFHo\_rotating}    \\ 

\hline

$19$ & SFHo & 0 & monopole & yes  & yes & 1.54 & 5.66 & 0.025 & \texttt{M19\_SFHo}   \\ 

\hline
\end{tabular}
\end{table*}

The 2D core-collapse supernova simulations analyzed in our study were
performed with the neutrino-radiation-hydrodynamics code 
\textsc{Fornax} (\citealt{skinner:16,burrows:16,vartanyan:18}, Skinner et al. 2018, in prep.).
\textsc{Fornax} solves the hydrodynamic equations using a 
directionally-unsplit Godunov-type finite-volume scheme in 
spherical coordinates, with the
HLLC approximate Riemann solver \citep{toro:94}. 
The majority of simulations presented here use a monopole
approximation for the approximate general-relativistic (GR)
gravitational potential, following Case A
of \citet{marek:06}. Some simulations were performed with a multipole gravity
solver \citep{mueller:95}, where we set the maximum spherical harmonic
order equal to twelve. \textsc{Fornax} offers a possibility to include
rotation in 2D, which is used in one of our simulations.

In \textsc{Fornax}, we distinguish three species of neutrino, i.~e., 
electron neutrinos $\nu_e$, anti-electron neutrinos $\bar{\nu}_e$,
and heavy lepton neutrinos ``$\nu_{\mu}$'', with the latter including
$\nu_{\mu}$, $\nu_{\tau}$, $\bar{\nu}_{\mu}$ and $\bar{\nu}_{\tau}$
taken together \citep{burrows:16}. The transport of neutrinos is followed using an explicit
Godunov characteristic method, with the HLLE approximate Riemann
solver \citep{einfeldt:88}, modified as in \citet{audit:02} 
and \citet{oconnor:15} to reduce the numerical dissipation in 
the diffusive limit. We use an M1 tensor closure for the $2^{\rm nd}$ and
$3^{\rm rd}$ moments of the radiation fields 
\citep{vaytet:11,shibata:11,murchikova:17}. The neutrino energy
is dicretized in twenty groups, varying logarithmically in the range 
$1-300\,{\rm MeV}$ for the electron neutrinos 
and $1-100\,{\rm MeV}$ for the other neutrino species.

We follow the prescription for the neutrino-matter interactions outlined
in \citet{burrows:06}. For more details about the neutrino microphysics
implemented in \textsc{Fornax} see \citet{burrows:16} and \citet{radice:17} and
references therein. We include
the effects of many-body corrections to the axial-vector term in the
neutrino-nucleon scattering rate, as described in~\citet{horowitz:17}.
One of our models was simulated without the many-body correction in order
to distinguish its influence on the GW signal.

In our models, we use three different equations of state (EOS), namely,
the SFHo EOS \citep{steiner:13}, the Lattimer-Swesty EOS with nuclear 
incompressibility parameter $220\,{\rm MeV}$ \citep{lattimer:91}, and the 
DD2 EOS \citep{fischer:14,banik:14}. 

In our simulations,
we use the progenitor models obtained with the stellar evolution code KEPLER
by \citet{sukhbold:16} with the ZAMS masses of $10$, $13$, and $19\,M_{\odot}$.
Our radial grid consists of 678 points for the $10\,M_{\odot}$ model and
608 points for the $13$ and $19\,M_{\odot}$ models, spaced evenly with $\Delta r =0.5\,{\rm km}$
for $r\lesssim10\,{\rm km}$ and logarithmically for $r\gtrsim 100\,{\rm km}$,
smoothly transitioning in between. The outer boundary is placed at 20,000 km. 
The angular resolution smoothly varies between $\approx 0.95^{\circ}$
at the poles and $\approx 0.65^{\circ}$ at the equator in 256 zones.
To avoid the overly restrictive Courant conditions close to the coordinate center, 
the angular resolution decreases in the innermost radial zones, 
representing a so-called dendritic grid (Skinner et al. 2018, in prep.).

To extract the GW signal measured by a distant observer we employ the 
standard formula for the trace-free quadrupole moment of the source in the
slow-motion approximation \citep{finn:90,murphy:09}:
\begin{equation}
\rqm_{jk}=\int \rho  \left(x^j x^k -\frac{1}{3}\delta^{jk}x_ix^i\right) d^3x\ .
\end{equation}
For axisymmetric sources, this
has only one independent component along the symmetry axis, $\rqm_{zz}$. 
In spherical coordinates, the time
derivative of this component can be rewritten in terms of the fluid velocity $v_i$
as (Equation (38) of \citealt{finn:90}, corrected in \citealt{murphy:09}):
\begin{eqnarray}
&&\frac{d}{dt}\rqm_{zz} = \frac{8\pi}{3}\int_{-1}^{1}d\cos\theta \int_{r_1}^{r_2}dr r^3 
\rho \times \nonumber \\ && \qquad \qquad \left[ P_2(\cos\theta)v_r + 
\frac{1}{2}\frac{\partial}{\partial\theta}P_2(\cos\theta)v_{\theta}\right]\ ,
\end{eqnarray}
where $P_2(\cos\theta)$ is the second Legendre polynomial.
After that, the axisymmetric GW strain can be computed as 
\begin{equation}
h_+ =\frac{3}{2}\frac{G}{D c^4}\sin^2\theta'\frac{d^2}{dt^2}\rqm_{zz}\ ,
\end{equation}
where $D$ is the distance to the source and $\theta'$ is the angle between 
the symmetry axis and the line of sight of the observer 
(henceforth, we assume $\sin^2\theta'=1$).
Following~\citet{murphy:09}, we compute the total energy emitted in GWs as
\begin{equation}
E_{GW}=\frac{3}{10}\frac{G}{c^5}\int_0^t\left(\frac{d^3}{dt^3}\rqm_{zz}\right)^2 dt\ ,
\end{equation}
and compute the spectrogram of this energy by means of the short-time
Fourier transform (STFT)
\begin{equation}
\frac{dE^*_{GW}}{df}(f,\tau) = \frac{3}{5}\frac{G}{c^5}\left(2\pi f\right)^2 
\left|\tilde{S}(f,\tau)\right|^2\ ,
\end{equation}
where
\begin{equation}
\tilde{S}(f,\tau)=\int_{-\infty}^{\infty}A(t)H(t-\tau)e^{-2\pi i f t} dt\ ,
\end{equation}
$A\equiv\frac{d^2}{dt^2}\rqm_{zz}$, and $H(t-\tau)$ is the Hann window
function with the time offset $\tau$. The sampling frequency of the GW strain
output in our simulations is $16,384\,{\rm Hz}$, and we use the window size of $40\,{\rm ms}$,
when performing the STFT. The high sampling frequency is necessary to avoid the
aliasing in GW spectrograms, seen in some of the early studies.

In addition to the matter motion, we compute the GW signal associated
with the neutrino emission, first recognized by \citet{epstein:78} 
\citep[see more in][]{thorne:92,burrows:96,mueller:97}. We use Eq.~(24)
from \citet{mueller:97} for the transverse-traceless part of the gravitational 
strain from neutrinos, $h_{ij}^{\rm TT}$, which we provide here for completness
\citep[see also][]{yakunin:15}:
\begin{equation}
h_{ij}^{\rm TT} = \frac{4G}{c^4D}\int_{-\infty}^{t-D/c}dt'\int_{4\pi}
d\Omega'\frac{(n_i n_j)^{\rm TT}}{1-\cos\Theta}\frac{dL_{\nu}(\mathbf{\Omega}',t')}{d\Omega'}\ ,
\end{equation}
where $\Theta$ is the angle between the direction towards the
observer and the direction $\mathbf{\Omega}'$ of the radiation emission, and
$dL_{\nu}(\mathbf{\Omega},t)/d\Omega$ is the direction-dependent neutrino
luminosity, defined as the energy radiated at time $t$ per unit of time and per 
unit of solid angle into direction $\mathbf{\Omega}$. Here, $n_i$ is the unit vector
in the direction of neutrino emission whose components are given with respect to
the observer's frame.

Table~\ref{tab:models} summarizes the set of simulations analyzed in the current study.
Some of these simulations were published before in \citet{radice:17}, while many of them
are described in more detail in \citet{vartanyan:18}. These
models are collected here to summarize and encompass the key dependences of the GW signal on the
intrinsic parameters of the progenitor, such as its mass and angular velocity, and
on the physical assumptions used in the code, such as the EOS, inclusion of the
many-body corrections, and the gravity solver. In Table~\ref{tab:models}, the total energy emitted in GWs,
$E_{\rm GW}$, is calculated up to the point where the simulation ends. We compute
this energy separately for the matter and the neutrino components of the GW signal. 
The GW energy associated with the anisotropic neutrino emission 
constitutes a few percent of the total energy emitted in GWs. For the rotating model,
the initial cylindrical rotational angular frequency depends on the radial coordinate as
$\Omega_0\left(1+(r/A)^2\right)^{-1}$, where $A=10,000\,{\rm km}$.

\section{Results}

In this section, we describe the main results of our study. We 
present an example of the GW signal from one of our numerical
models and discuss its key features, which
are common for all our models. After that, we address the physical nature 
of the main components of the GW signal with the help of linear perturbation
analysis. In particular, we demonstrate that the strongest component of the GW signal 
is associated with the fundamental (f) $l=2$ mode of the PNS.
Finally, we show the dependence on the GW signal on the 
progenitor mass, EOS, and example variation in the neutrino microphysics. In addition, we present
the GW signal from a rotating progenitor model, obtained with full neutrino physics
in 2D and calculated to $\sim$1 second after bounce.

\begin{figure}
  \centering
  \includegraphics[width=0.462\textwidth]{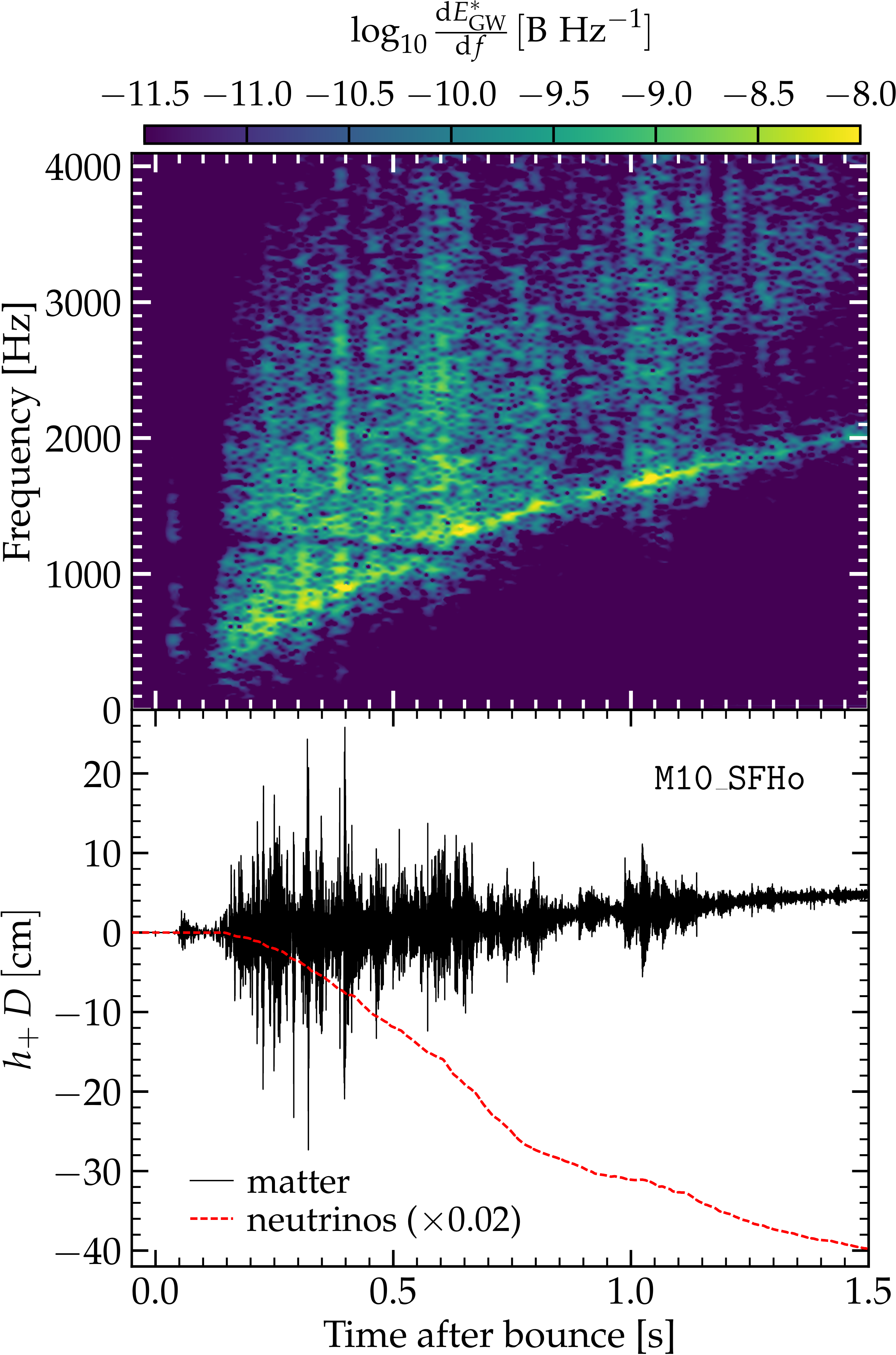}
  \caption{Spectrogram (top) and the corresponding waveform (bottom) of 
  the GW signal from the model \texttt{M10\_SFHo}.} 
  \label{fig:title}
\end{figure}

\subsection{Structure of the GW signal from CCSNe: the
\texttt{M10\_SFHo} model}
\label{example}

It is common in the literature to distinguish four components of the
GW signal from CCSNe, namely, the prompt convection signal, 
the quiescent phase, the neutrino convection/SASI 
driven phase, and the explosion phase \citep[see, e.g.,][]{murphy:09,mueller:13,yakunin:17}.
Here, we demonstrate these components using as an example our
non-rotating $M_{\rm ZAMS}=10\,M_{\odot}$ model (\texttt{M10\_SFHo}) with the SFHo
EOS, including the many-body corrections to the neutrino-nucleon
scattering cross section \citep{horowitz:17}.
This model starts to explode at $\sim$400$-$600$\,{\rm ms}$ after bounce, 
which allows us to
address both pre- and post-explosion regimes.

Figure~\ref{fig:title} shows the GW spectrogram and the strain
times distance, $h_+D$, for the model \texttt{M10\_SFHo}. 
The GW strain is
shown both for the matter (black) and the neutrino (red) contributions.
The amplitude of the GW signal due to the anisotropic neutrino emission
is about two orders of magnitude larger than the amplitude of the signal
related to mass motions. Its characteristic frequency, however,
does not exceed several tens of Hz. In this study, we do not focus on the
GW signal due to neutrinos, and the spectrogram in the top panel of Figure~\ref{fig:title}
takes into account only the matter contribution. 

As in previous studies \citep[see, for example,][]{marek:09,murphy:09,mueller:13,yakunin:10,yakunin:17}, 
we see the early signal associated with the prompt PNS convection in the first $\sim$50$\,{\rm ms}$ after
bounce. The duration and strength of this signal depend upon the progenitor mass
and EOS, but this component is generally weak compared to the other, more
dominant features in the spectrogram. The only exception is the rotating $13\,M_{\odot}$ model, 
which manifests a very energetic prompt convection signal and will be shown later in 
Section~\ref{rotation}. This is expected based on previous work devoted
to the GWs from rotating core collapse \citep{dimmelmeier:08,abdikamalov:14,richers:17,torres:17}.
The prompt convection signal is followed by a short, 
$\sim$50$\,{\rm ms}$, quiescent phase, in agreement with previous
results \citep{marek:09,murphy:09,mueller:13,yakunin:17}.

\begin{figure}
  \centering
  \includegraphics[width=0.49\textwidth]{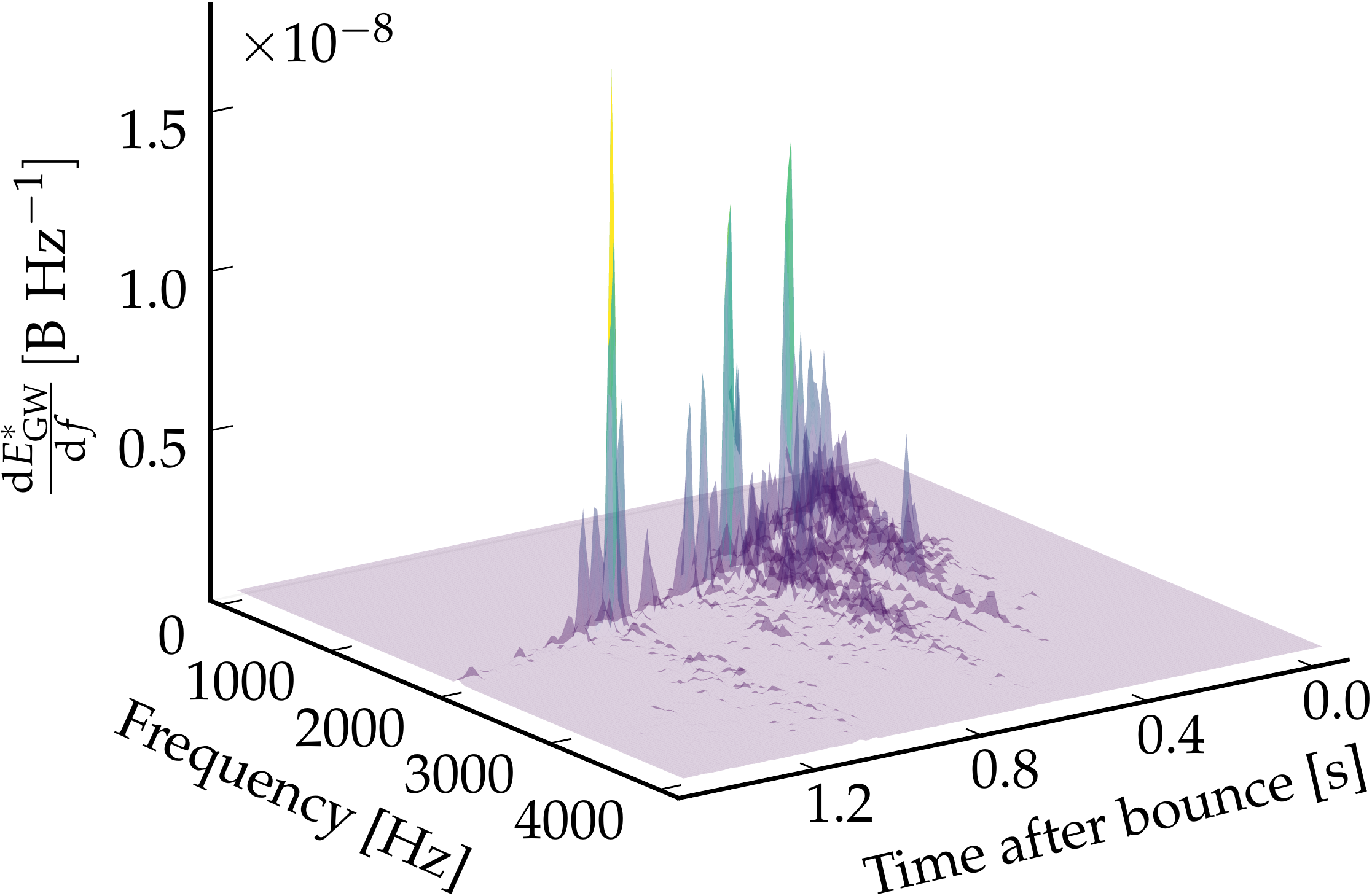}
  \caption{Linear 3D representation of the GW spectrogram from the 
  model \texttt{M10\_SFHo}.} 
  \label{fig:3d10}
\end{figure}

The dominant part of the signal lasts from $\sim$150$\,{\rm ms}$ after core bounce
until the end of the simulation, with the frequency growing from $\sim$300 to $\sim$2000$\,{\rm Hz}$.
Despite the high-frequency noise, most of the energy is concentrated along
a relatively thin stripe, as can be seen from the linear 3D visualization of the
spectrogram in Figure~\ref{fig:3d10}. Some of the earlier work predicted the abrupt reduction in the
high frequency signal at the onset of explosion due to the cessation of 
down-flowing plume excitation of the inner core \citep{murphy:09,yakunin:15}.
However, as was shown in \citet{mueller:13}, the high frequency signal may persist
for a certain time before this happens, and we see the same in our model. As in \citet{mueller:13}, 
the post-explosion signal from our model \texttt{M10\_SFHo} consists of distinct 
`bursts' of emission, presumably caused by the continuing accretion episodes.
For another exploding model in our study
($19\,M_{\odot}$), the post-explosion signal stays strong until the end of the
simulation at $\sim$1.5$\,{\rm s}$ after bounce, without decaying in energy 
(see more in Section~\ref{dependences}). The explosion is 
marked by the offset of $h_+D$ from zero, which indicates that the shock is not
spherical (the prolate explosion shifts the strain up, while the oblate explosion
shifts it down; see \citealt{murphy:09,mueller:13,yakunin:15}).

\begin{figure}
  \centering
  \includegraphics[width=0.48\textwidth]{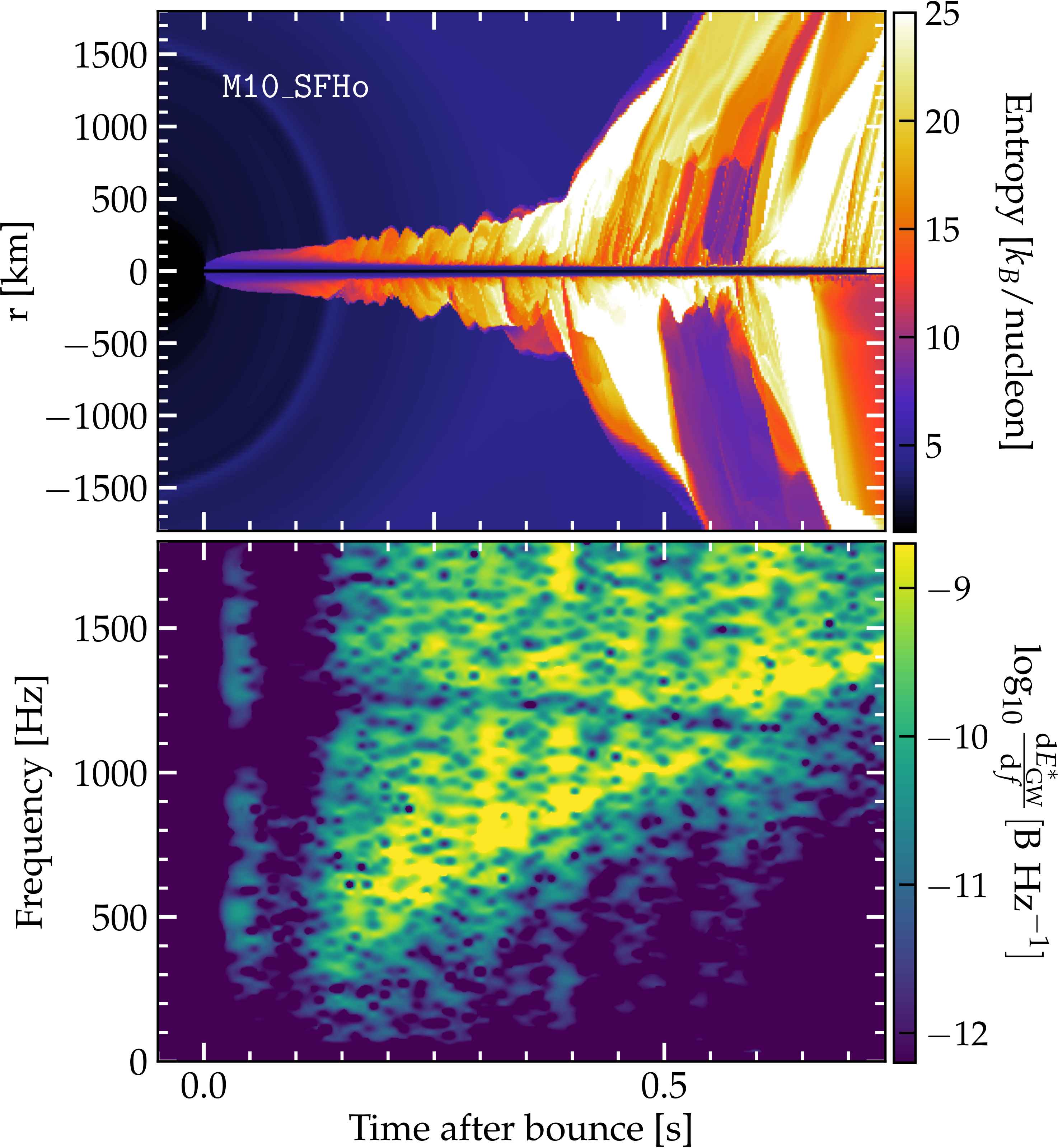}
  \caption{Top panel: Entropy along the north and south polar axis as a function of time
  for \texttt{M10\_SFHo}. Bottom panel: Zoomed in early part of the GW spectrogram for this
  model. We associate the weak power excess at low frequencies between $100$ and 
  $400\,{\rm ms}$ after bounce with the shock oscillations seen in the top panel.} 
  \label{fig:zoomed}
\end{figure}

A number of recent works \citep{cedraduran:13,kuroda:16,kuroda:17,pan:17,andresen:17} 
pointed to a separate GW feature associated
with the SASI \citep[see more about the SASI phenomenon, for example, in][]{blondin:03,foglizzo:07}. This signal
is expected to reside at lower frequency, typically $100-200\,{\rm Hz}$, and coincides in time with the periods of
enhanced shock oscillations. To test this regime in model \texttt{M10\_SFHo}, we plot
its entropy along the polar axis in the top panel of Figure~\ref{fig:zoomed}. The plot
shows that the shock oscillates mildly in the period $100-400\,{\rm ms}$ after bounce
(these oscillations, though, are not as vigorous as typically 
seen when the SASI is identified) and before
the explosion sets in. The early part of the GW spectrogram, plotted in the bottom panel
of Figure~\ref{fig:zoomed}, indeed shows some power excess at low frequencies in this 
period, and we associate it with the oscillations of the shock, 
but this signal is very weak compared to the higher frequency signal from
the same model. In the rest of this paper, we concentrate on the dominant part
of the GW signal at higher frequencies.

One curious feature seen in all our models is a `gap' crossing the GW spectrogram at
$\sim$1300$\,{\rm Hz}$. We checked the dependence of this feature, which at first glance
looked like a numerical artifact, on simulation parameters, such as timestep, 
resolution and output frequency. For example, the dependence on the GW signal
on the grid resolution for the model \texttt{M10\_LS220\_no\_manybody} 
is given in Appendix~\ref{Ap0}. 
The `gap' persisted at exactly the same location
for all combinations of numerical parameters we considered for a given model, varying 
only slightly between the different models. We do not exclude the possibility that
the `gap' is physical and attempt to explain it in part with the trapped g-mode of
the PNS inner core in the next subsection.

\subsection{Analytical explanation of the key features of GW signal from the
\texttt{M10\_SFHo} model}
\label{analysis}

In this subsection we focus on explaining the dominant features of the
GW spectrogram, using the model \texttt{M10\_SFHo} as an example
(shown in Figure~\ref{fig:title}), by means of a linear perturbation analysis.

The system of equations we solve combines the linearized equations
of general-relativistic hydrodynamics in a spherically-symmetric
conformally-flat background metric \citep{banyuls:97,torres:17}
together with the Poisson equation. It can be summarized in the form
\begin{eqnarray}
&&\label{fineq1t}\partial_r \eta_r + \left[\frac{2}{r}+\frac{1}{\Gamma_1}\frac{\partial_r P}{P} +
6\frac{\partial_r \psi}{\psi}\right]\eta_r \nonumber \\
&& \qquad\qquad\quad + \frac{\psi^4}{\alpha^2 c_s^2}
\left(\sigma^2-\mathcal{L}^2\right)\eta_{\bot}
-\frac{1}{\alpha c_s^2}\delta\hat{\alpha} = 0\ , \\ 
&&\label{fineq2t}\partial_r \eta_{\bot} - \left(1-\frac{\mathcal{N}^2}{\sigma^2}\right)\eta_r 
+ \left[\partial_r\ln q-\tilde{G}\left(1+\frac{1}{c_s^2}\right)\right]\eta_{\bot} \nonumber \\
&& \qquad\qquad\qquad\qquad\qquad\qquad\quad
-\frac{1}{\alpha \tilde{G}}\frac{\mathcal{N}^2}{\sigma^2}\delta\hat{\alpha} = 0\ , \\
&&\label{fineq3t}\partial_r f_{\alpha} +\frac{2}{r}f_{\alpha} + 4\pi \left[\partial_r\rho - 
\frac{\rho}{P\Gamma_1}\partial_r P\right]\eta_{r}  \nonumber \\  
&& \quad - \frac{4\pi\rho}{P\Gamma_1}
q\sigma^2\eta_{\bot} +\left[\frac{4\pi\rho^2 h}{P\Gamma_1\alpha}-
\frac{1}{\alpha}\frac{l(l+1)}{r^2}\right]\delta\hat{\alpha} = 0\ , 
\end{eqnarray}
and
\begin{equation}
\label{fineq4t} \partial_r \delta\hat{\alpha} = f_{\alpha}\alpha - 
\tilde{G}\delta\hat{\alpha}\ .
\end{equation}
Here, scalar functions $\eta_r = \eta_r(r)$ and $\eta_{\bot} = \eta_{\bot}(r)$
represent the amplitudes of the decomposition of radial ($\xi^r$) and polar ($\xi^{\theta}$)
Lagrangian displacements of a fluid element with respect to its eqilibrium position
in terms of spherical harmonics:
\begin{eqnarray}
&&\xi^r=\eta_r Y_{lm}e^{-i\sigma t}\ , \nonumber \\
&&\xi^{\theta}=\eta_{\bot}\frac{1}{r^2}\partial_{\theta}Y_{lm}e^{-i\sigma t}\ ,
\end{eqnarray}
where $\sigma$ is the mode frequency. The scalar function 
$\delta\hat{\alpha} = \delta\hat{\alpha}(r)$ is the amplitude of the 
lapse function perturbation
\begin{equation}
\delta \alpha = \delta \hat{\alpha}\,Y_{lm}e^{-i\sigma t}\ ,
\end{equation}
and we define $f_{\alpha}=\partial_r(\delta\hat{\alpha}/\alpha)$.
The conformal factor of the metric, $\psi$, is equal to 1 in our numerical
setup. The details of the derivation of Eqs.~(\ref{fineq1t})-(\ref{fineq4t}) are
given in Appendix~\ref{Ap1}. In the limit $\delta\alpha=0$
(the Cowling approximation), Eqs.~(\ref{fineq1t})-(\ref{fineq2t}) coincide with 
Eqs.~(31)-(32) of \citet{torres:17}.

In Eqs.~(\ref{fineq1t})-(\ref{fineq4t}), $P$ is the pressure, 
$\rho$ is the rest-mass density of
the matter, $h$ is the specific enthalpy, $c_s$ is the relativistic speed
of sound, $\Gamma_1$ is the adiabatic index, $\tilde{G}\equiv-\partial_r\ln\alpha$ 
is the radial component of the gravitational acceleration,
$q\equiv \rho h\alpha^{-2}\psi^4$, $\mathcal{N}$ is the relativistic 
Brunt-V\"{a}is\"{a}l\"{a} frequency,
which in our case is equal to \citep[see also][]{mueller:13}
\begin{equation}
\mathcal{N}^2 = \frac{\alpha \partial_r \alpha}{\psi^4} \left(\frac{1}{\Gamma_1}\frac{\partial_r P}{P} 
- \frac{\partial_r e}{\rho h} \right)\ ,
\end{equation}
and $\mathcal{L}$ is the relativistic Lamb frequency
\begin{equation}
\mathcal{L}^2=\frac{\alpha^2}{\psi^4} c_s^2 \frac{l(l+1)}{r^2}\ .
\end{equation}
These quantities describe the spherically 
symmetric equilibrium background configuration, which we find
by averaging the hydrodynamical output of our 2D simulations
over polar angle\footnote{To compute the lapse
function of the equilibrium background configuration from the 
simulation output, we use the formula $\alpha=\exp\left(\Phi_{\rm eff}/c^2\right)$,
where $\Phi_{\rm eff}$ is the approximate relativistic gravitational potential
(\citealt{marek:06}; Case A).}.

\begin{figure}
  \centering
  \includegraphics[width=0.49\textwidth]{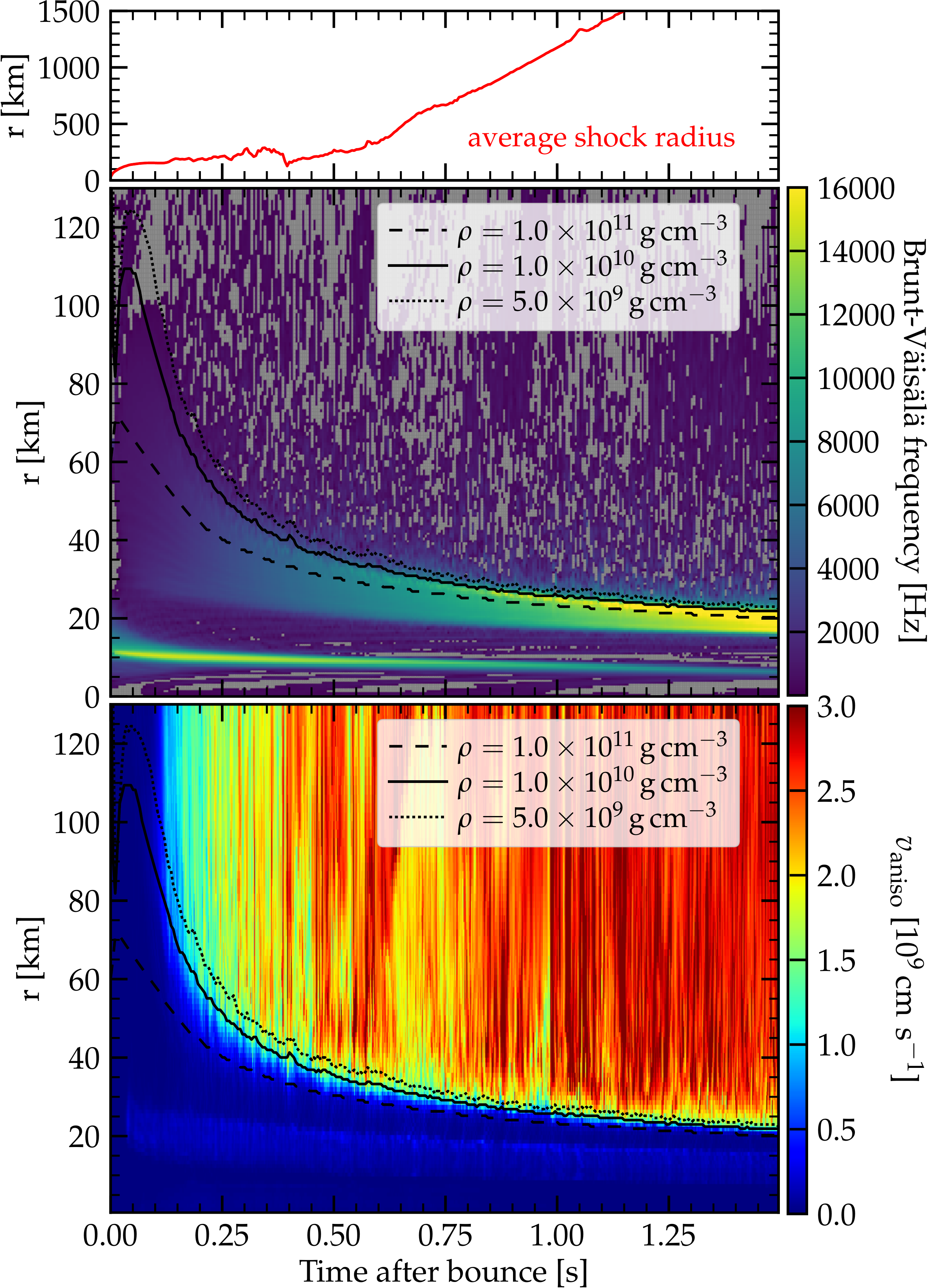}
  \caption{Top panel: Average shock radius of the model \texttt{M10\_SFHo} 
  as a function of time. Middle panel: Brunt-V\"{a}is\"{a}l\"{a} frequency 
  ($\mathcal{N}$) of the averaged profile of this model
  as a function of time and radial coordinate. Gray color corresponds to
  negative values of $\mathcal{N}^2$, marking the regions that are 
  convectively unstable. Bottom panel: Anisotropic velocity
  of the model \texttt{M10\_SFHo} as a function of time and radial coordinate.
  } 
  \label{fig:BV}
\end{figure}

Figure~\ref{fig:BV} shows the Brunt-V\"{a}is\"{a}l\"{a} frequency for the 
averaged profile of the \texttt{M10\_SFHo} model (middle panel). Black lines show the radial
coordinates where the density is equal to $5.0\times10^{9}\,{\rm g}\,{\rm cm}^{-3}$,
$10^{10}\,{\rm g}\,{\rm cm}^{-3}$, and $10^{11}\,{\rm g}\,{\rm cm}^{-3}$, with the 
latter density surface commonly used as a definition of the PNS boundary.
Colored gray are the regions where the Brunt-V\"{a}is\"{a}l\"{a} frequency
is imaginary ($\mathcal{N}^2<0$), which means they are convectively
unstable. To further emphasize the convection, in the
bottom panel of Figure~\ref{fig:BV} we plot the anisotopic velocity
defined as \citep{takiwaki:12,pan:17}
\begin{equation}
v_{\rm aniso} = \sqrt{\frac{\left\langle\rho\left[\left(v_r-\langle v_r\rangle_{4\pi}\right)^2
+v_{\theta}^2\right]\right\rangle_{4\pi}}{\langle\rho\rangle_{4\pi}}}\ ,
\end{equation}
where $\langle\rangle_{4\pi}$ denotes spherical averaging. From
this plot, one can see a convective layer inside the PNS, 
between $\sim$10 and $\sim$20$\,{\rm km}$, but the convective
velocities there are much smaller than the ones above the PNS
surface. The outer convective zone 
between the PNS boundary radius and the shock radius 
(shown in the top panel of Figure~\ref{fig:BV})
is recognized as the main driving region for the GW
signal \citep{murphy:09}.
The middle panel of Figure~\ref{fig:BV} shows the imaginary
Brunt-V\"{a}is\"{a}l\"{a} frequency in the center of the PNS, however,
the bottom panel shows no convection in that region.

\begin{figure*}
  \centering
  \includegraphics[width=0.99\textwidth]{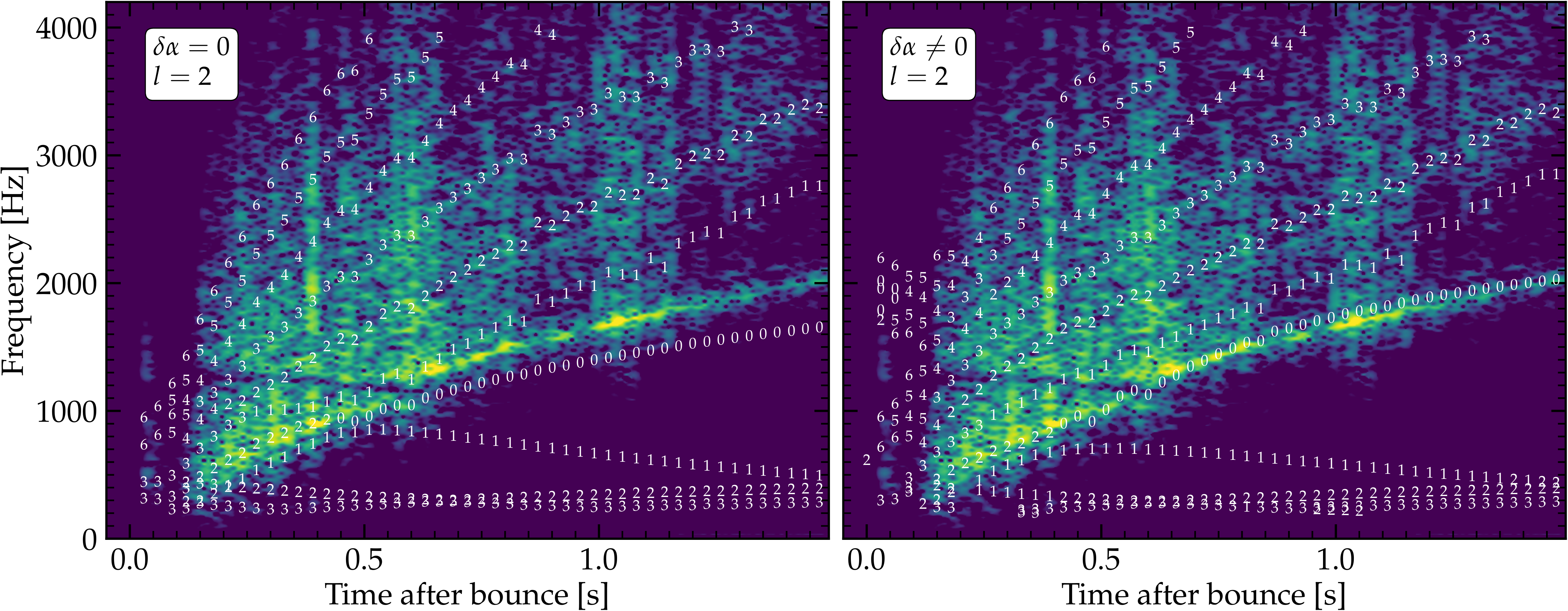}
  \caption{Eigenfrequencies $\sigma/2\pi$ of the $l=2$ modes compared to the GW spectrogram
  from model \texttt{M10\_SFHo}. Each digit represents the number of nodes in the 
  corresponding mode. The left panel shows the results obtained using the
  Cowling approximation, while the right panel shows the solution of the full system of 
  Eqs.~(\ref{fineq1t})-(\ref{fineq4t}). 
  In the right panel, the dominant feature of
  the spectrogram is well described by the fundamental (0 radial nodes) mode
  starting from $\sim$400$\,{\rm ms}$ after bounce.} 
  \label{fig:l2}
\end{figure*}
\begin{figure}
  \centering
  \includegraphics[width=0.475\textwidth]{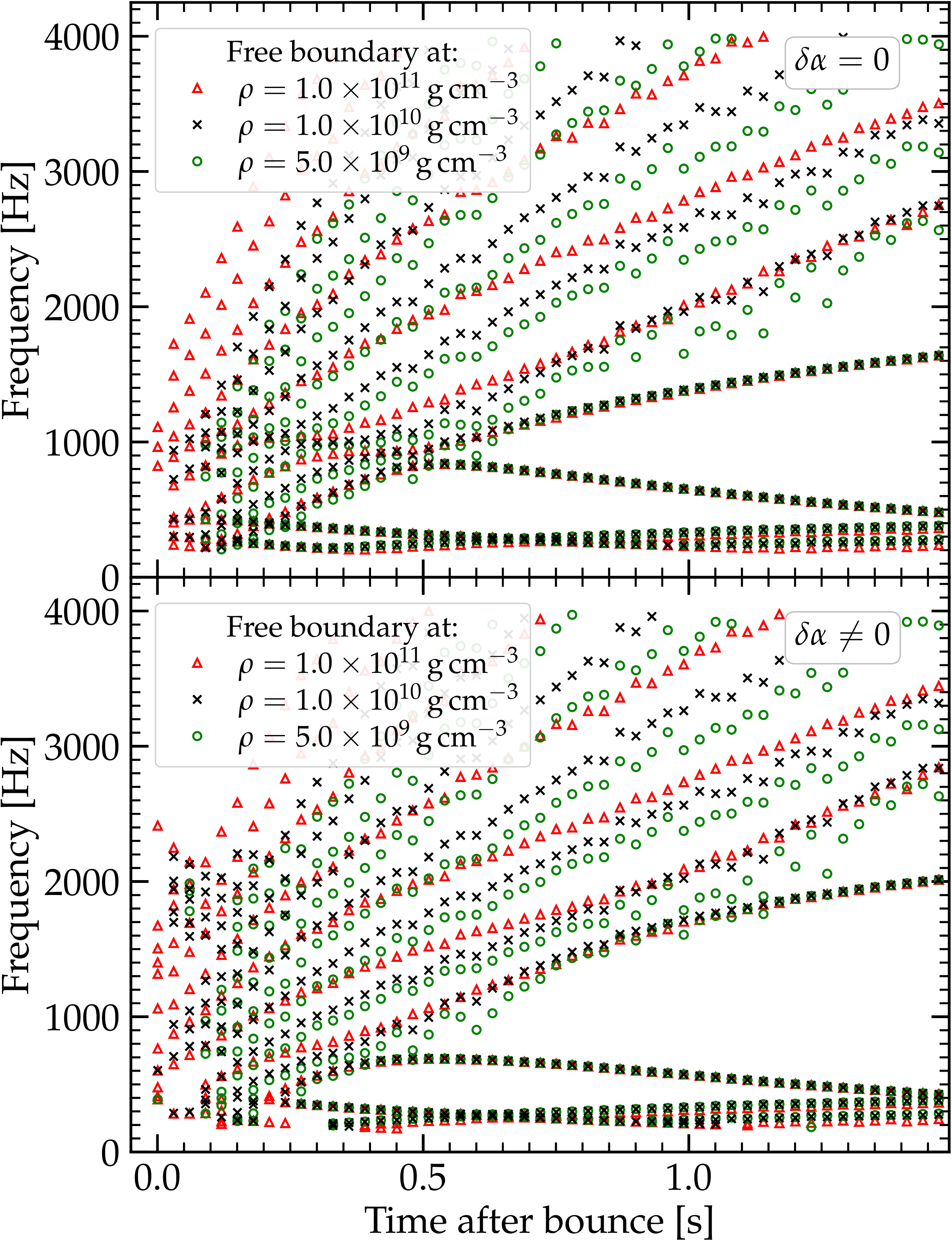}
  \caption{Dependence of the derived eigenfrequencies on the position of the
  outer boundary in our analysis. This plot demonstrates that the frequencies of 
  p-modes are only approximately captured by our calculations. At the same time, the frequencies of the f-mode
  and the low order g-modes are almost insensitive to the position of the outer
  boundary, which demonstrates the robustness of our main result, i.e., the
  association between the dominant GW feature and the fundamental (f) $l=2$ PNS mode.} 
  \label{fig:l2_bound}
\end{figure}

To solve Eqs.~(\ref{fineq1t})-(\ref{fineq4t}), we place the outer boundary
condition at the radial coordinate where $\rho=10^{10}\,{\rm g}\,{\rm cm}^{-3}$
(solid black line in Figure~\ref{fig:BV}). There, we impose the condition 
$\Delta P =0$ on the Lagrangian perturbation of the pressure, which
physically corresponds to a free surface of the 
PNS \citep[see, for example,][]{reisenegger:92}.
Mathematically, this boundary condition can be written as
\begin{equation}
\label{outer}
q\sigma^2\eta_{\bot} - \frac{\rho h}{\alpha}\delta\hat{\alpha} + \partial_r P\eta_r = 0\ .
\end{equation}
Our treatment of the outer boundary condition is, therefore, 
different from the one in \citet{torres:17},
where $\eta_r=0$ at the shock position is imposed instead.
At the innermost point, we impose a small radial 
displacement, use the regularity condition \citep{reisenegger:92}
\begin{equation}
\eta_r|_{r=0} = l \eta_{\bot}|_{r=0}\ ,
\end{equation}
and assume $\delta\hat{\alpha}|_{r=0}=f_{\alpha}|_{r=0}=0$. 
As in \citet{torres:17}, we apply a trapezoidal rule to discretize
the radial derivatives in Eqs.~(\ref{fineq1t})-(\ref{fineq4t}). Starting
from the innermost point, we integrate the equations outwards,
inverting a $4\times4$ matrix of coefficients at every step. We use the
bisection method to find the solutions satisfying Eq.~(\ref{outer})
at the outer boundary. The frequencies $\sigma/2\pi$ corresponding to
these solutions are the eigenfrequencies of our model.

Figure~\ref{fig:l2} shows the eigenfrequencies of $l=2$ 
(quadrupolar) modes overplotted on the GW spectrogram for the
model \texttt{M10\_SFHo}. Each eigenfrequency is represented
by a number of nodes in the corresponding mode, i.e., the number
of times the radial displacement function $\eta_r$ changes its sign
along the radial coordinate. To avoid crowding the numbers, we show
only the modes with the number of nodes $<\,$7 above the frequency
$700\,{\rm Hz}$ and $<\,$4 below that frequency.
Since the GW signal itself was obtained from 
the numerical simulations using a quadrupole formula \citep{finn:90},
we primarily focus on $l=2$ modes in this study. 
At the same time, we cannot 
exclude the case of non-linear coupling between the $l=2$ modes
and the modes of different $l$s, which can explain certain features
of the GW signal \citep[see, for example,][]{torres:17}. For the interested
reader, the $l=3$ and $l=4$ modes are shown in 
Appendix~\ref{Ap2}.

The left panel of Figure~\ref{fig:l2} shows the results for the modes obtained
under the Cowling approximation 
($\delta\hat{\alpha}=0$ and $f=0$ in Eqs.~(\ref{fineq1t})-(\ref{fineq4t})).
Starting from $\sim$0.4$\,{\rm s}$ after bounce, a 
fundamental mode (the f-mode, with zero radial nodes) can be clearly identified.
Above this mode, one can see p-modes (acoustic), for which the 
frequencies increase with the number of nodes, while below it there are
g-modes, for which the frequencies decrease with the number of nodes.
Before $\sim$0.4$\,{\rm s}$ after bounce, as in \citet{torres:17}, we see the
mixing and crossing between the different modes, during which they
change the number of nodes.

\begin{figure}
  \centering
  \includegraphics[width=0.475\textwidth]{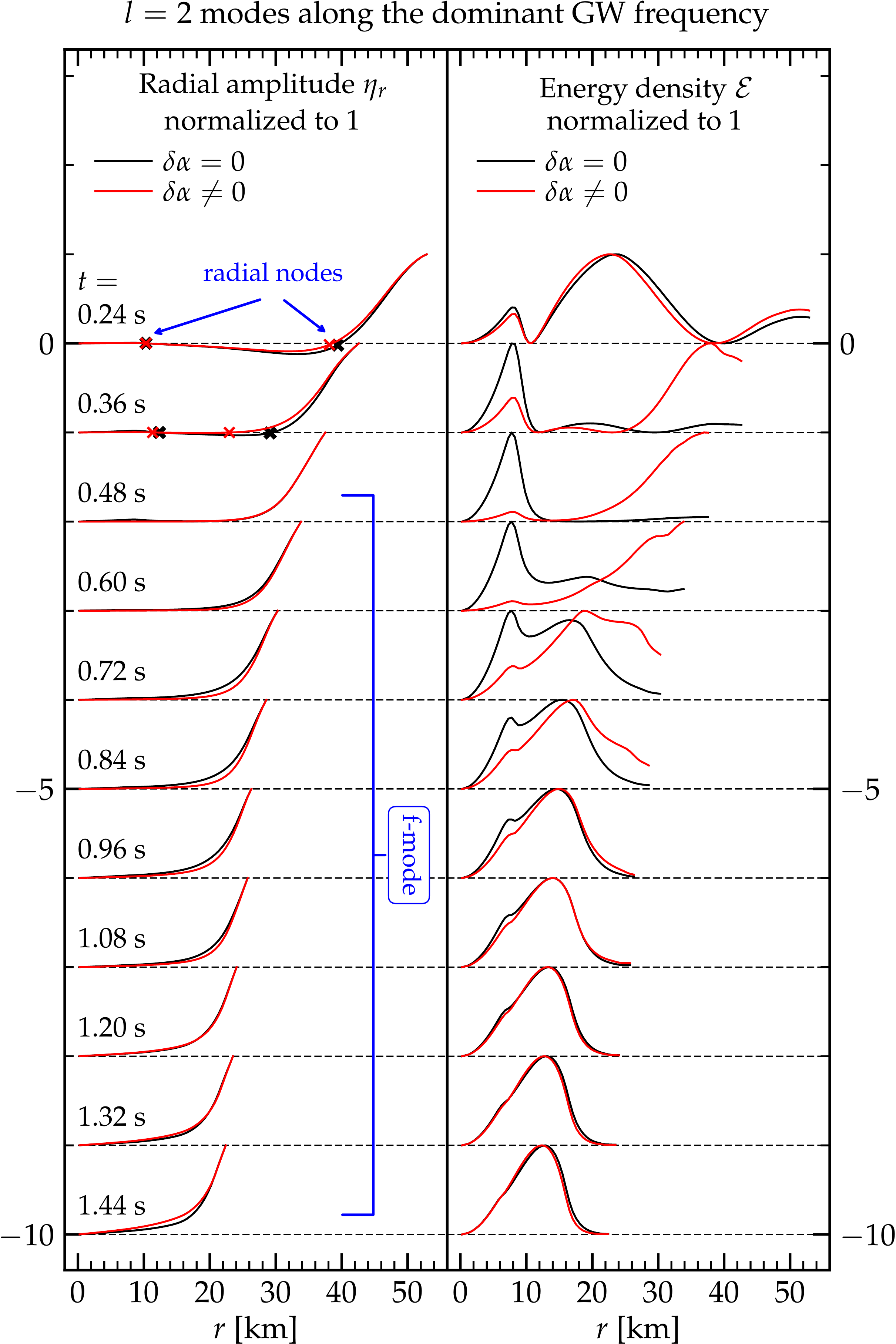}
  \caption{Normalized radial eigenfunction $\eta_r$ (left) and the associated energy
  density $\mathcal{E}$ (right) of the $l=2$ modes tracing the dominant component 
  of the GW signal as a function of radius for several
  subsequent times (the time is indicated along the left hand side of the plot). At the
  early times ($\sim$200$-$400$\,{\rm ms}$ after bounce), the dominant mode is a g-mode
  with 2 radial nodes, while starting from $\sim$400$\,{\rm ms}$ after bounce it is the f-mode.
  Crosses indicate the position of the radial nodes.
  Black color shows the results obtained using the
  Cowling approximation, while the red color shows the solution of the 
  system of Eqs.~(\ref{fineq1t})-(\ref{fineq4t})
  when $\alpha\neq0$.
  The eigenfunctions are terminated at the location of the outer boundary at each time.
  The overall shape of the eigenfunction is very similar between the $\alpha=0$ and
  $\alpha\neq0$ cases.} 
  \label{fig:eigen}
\end{figure}
\begin{figure}
  \centering
  \includegraphics[width=0.475\textwidth]{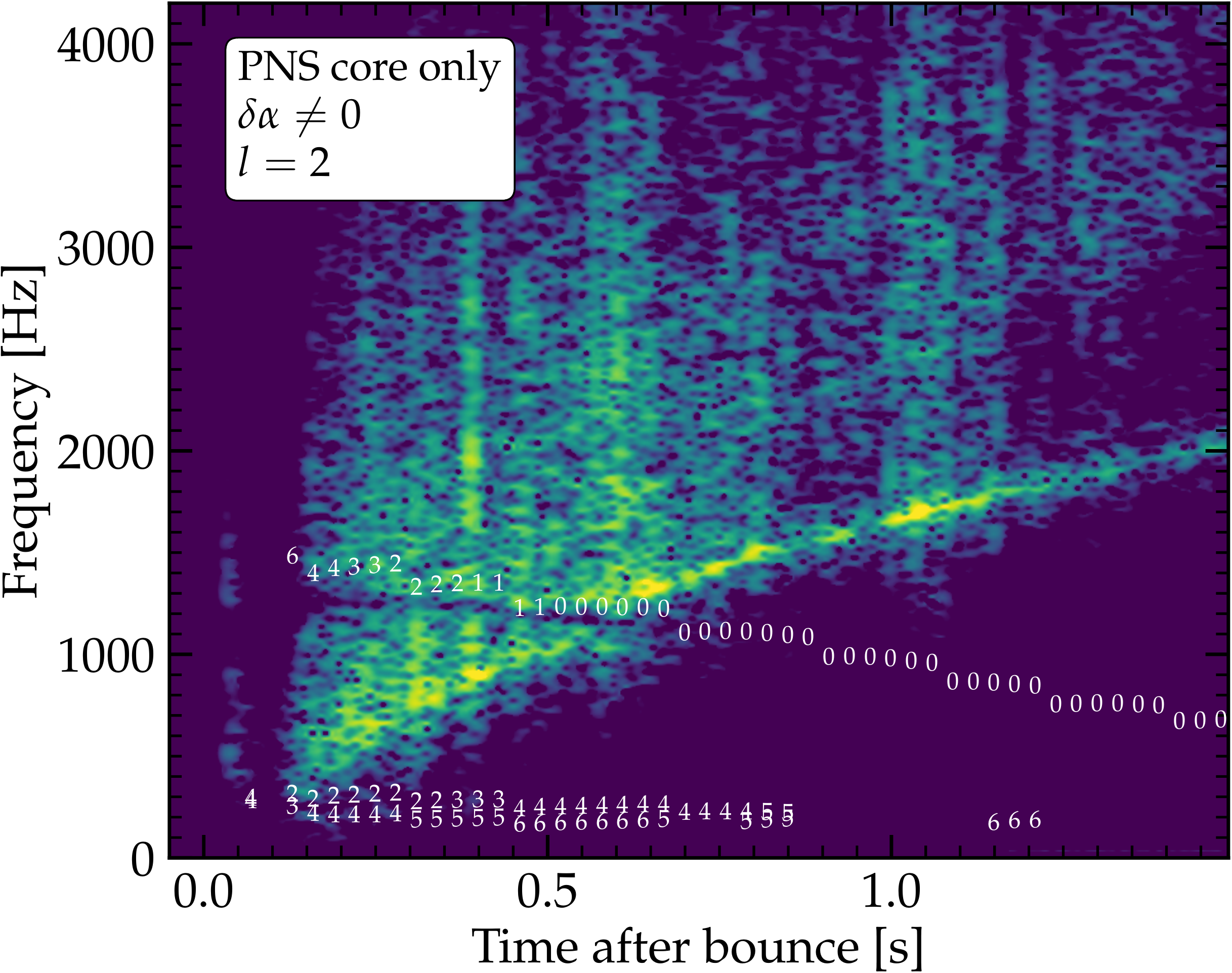}
  \caption{Eigenfrequencies $\sigma/2\pi$ of the $l=2$ modes found for the 
  PNS inner core only. Each digit represents the number of nodes in the 
  corresponding mode. Plotted are all modes with the number of nodes
  less than $7$. Step-like behavior of the eigenfrequencies is a result of the
  insufficient resolution of the PNS core in the simulations. Nevertheless, 
  the core eigenfrequencies lie close to the position of the `gap' in the GW
  spectrogram and roughly resemble its morphology. We speculate that the `gap'
  may appear as a result of interaction between the high-order p-modes
  and the trapped mode of the PNS inner core, e.g., by means of an
  avoided crossing.} 
  \label{fig:inner}
\end{figure}

The right panel of Figure~\ref{fig:l2} shows the full solution of 
Eqs.~(\ref{fineq1t})-(\ref{fineq4t}), for $\delta\hat{\alpha}\neq0$. 
As in the left panel, the
fundamental (f) mode clearly stands out after $\sim$0.4$\,{\rm s}$
post-bounce time, but in this case it agrees very well with the
strongest component of the GW radiation. This result is
expected from physical grounds, and it shows that
the Cowling approximation can indeed affect the analysis and
should be used with caution when interpreting the GW signal
from numerical simulations of CCSNe.
The p- and g-modes can be also identified 
in the left panel of Figure~\ref{fig:l2}.
Interestingly, the GW spectrogram from our simulations
shows almost no power below the f-mode, suggesting
that the higher order g-modes of the PNS are not excited.
Aside from the possible SASI and neutrino signal, which are expected to 
operate at the frequencies $\lesssim$100$\,{\rm Hz}$
\citep[see][]{kuroda:16,andresen:17}, there is no
other apparent mechanism that could fill this 
`excluded region' of the spectrogram.

Figure~\ref{fig:l2_bound} shows the dependence of the obtained
results on the position of the outer boundary, placed at the radial
coordinate where the density reaches a given value. We remind 
the reader that the three
choices of boundary density correspond to the three black lines in 
Figure~\ref{fig:BV} with the middle value, $\rho=10^{10}\,{\rm g}\,{\rm cm}^{-3}$, being
our default choice. From Figure~\ref{fig:l2_bound}, it is seen that
our approach does not let us capture the outer p-mode frequencies 
very accurately, because the result is very sensitive to the position of the outer
boundary. This is probably related to the fact that the p-modes
represent the sound waves propagating between the PNS surface
and the shock position, a region which is not taken into account
in our analysis. This, however, does not affect the qualitative conclusion
that the high frequency noise on the GW spectrogram above the 
dominant feature is at least partially associated with 
these modes. Another possible source of this noise
is the turbulent convection between the PNS and the shock front, which 
is chaotic and does not necessarily represent any simple eigenmode of 
the system.

At the same time, the frequency of the fundamental mode
in Figure~\ref{fig:l2_bound} is almost
insensitive to the position of the outer boundary, and the low-order g-modes
depend weakly on it. Importantly, this shows that
the dominant GW frequency is not just proportional to the 
Brunt-V\"{a}is\"{a}l\"{a} frequency at the surface of the PNS, as
was suggested in earlier work. 
Indeed, Figure~\ref{fig:BV} shows that the three black lines corresponding to the
different outer boundary locations pass through very different values of the 
Brunt-V\"{a}is\"{a}l\"{a} frequency. 
The fact that the fundamental quadrupolar eigenfrequency in Figure~\ref{fig:l2_bound}
is nearly independent on the position of the outer boundary tells us that the dominant frequency
of the GW signal is defined by the entire structure of the PNS, rather
than by its surface characteristics alone.

The left panel of Figure~\ref{fig:eigen} illustrates the time evolution of the radial
eigenfunction $\eta_r$ for the $l=2$ modes associated with the dominant frequency
of the GW signal. The eigenfunctions are normalized to 1 and plotted as a function
of radial coordinate from the innermost grid point up to the location of the outer boundary.
In Figure~\ref{fig:eigen}, they are shifted along the y-axis according to the time after bounce
at which they
are calculated (the time is indicated on the left side of the panel and directed downwards).
As we already mentioned, starting from $\sim$400$\,{\rm ms}$ after bounce and until the
end of the simulation the main signal is represented by the f-mode, which has the largest
amplitude at the PNS boundary surface and gradually decreases towards the center.
Before that, in the time interval between $\sim$200 and $\sim$400$\,{\rm ms}$, this
mode is smoothly connected to a g-mode having two radial nodes (see also the left
panel of Figure~\ref{fig:l2}). The right panel of Figure~\ref{fig:eigen} shows the 
energy density $\mathcal{E}$ defined as \citep{torres:17}
\begin{equation}
\mathcal{E} = \frac{\sigma^2}{8\pi} \rho \left[\eta_r^2+l(l+1)\frac{\eta_{\bot}^2}{r^2}\right]
\end{equation}
for the corresponding eigenfunctions of the left panel.
The figure shows that the shape of the fundamental eigenfunction
is very similar in the case of the Cowling approximation (black lines) and in the case when
$\delta\alpha\neq0$ (red lines). The energy density of the modes shows less
agreement. Note that the definition of $\mathcal{E}$ contains the mass density, which is
larger in the inner region than at the the surface of the PNS. Therefore, even a 
barely visible disagreement between the eigenfunctions in the inner region may lead to a 
large disagreement between the energy density distributions (see, for example, the $0.48\,{\rm s}$
snapshot in Figure~\ref{fig:eigen}).

Finally, in Figure~\ref{fig:inner} we attempt to address the nature of the `gap' 
seen in our spectrograms by performing the linear perturbation analysis of the
PNS inner core only. For that, we place the outer boundary at the inner maximum of the
Brunt-V\"{a}is\"{a}l\"{a} frequency, which roughly corresponds to the radial
coordinate of $10\,{\rm km}$ (see Figure~\ref{fig:BV}), and solve the system of
Eqs.~(\ref{fineq1t})-(\ref{fineq4t}) in that inner region, using the boundary condition~(\ref{outer}). 
This approach is not strictly accurate, but it gives us an idea about the eigenfrequencies
of the inner core. In Figure~\ref{fig:inner}, the digits show the number
of radial nodes in the corresponding modes. 
All modes with the number of nodes less than $7$ are shown, without
selection.
The resolution of the PNS core
in our simulations does not exceed a few tens of grid points, which leads to the
spurious nodes and the step-like behavior of the eigenfrequencies 
(for the same reason, it does not make sense to plot the modes with a larger
number of nodes). Nevertheless, a part of
the core eigenfrequencies lies very close to the `gap' position in the spectrogram and
roughly reproduces its morphology. 

We speculate that the `gap' may be the result
of interaction between the trapped PNS core mode and the
other (p- or f-) modes of the system, probably by means
of an avoided crossing \citep{christensen:81,yoshida:01,stergioulas:03}. One of the
simplest examples of the avoided crossing phenomenon is the case of two 
coupled classical oscillators, where the eigenfrequencies demonstrate
characteristic splitting in the strong coupling regime \citep{novotny:10}. Analogously,
one may view the PNS as a coupled system of the inner core and the outer
convectively stable shell, mediated by the inner PNS convection region \citep{dessart:06}.
In this picture, the modes of the inner core may repel the modes of
the shell, leading to an empty region in the frequency space with the width 
related to the strength of the coupling between the two. At the same time, 
the inner mode itself is most likely not excited, because
it is shielded from the down-falling plumes of the postshock convection region by
the PNS surface (though, the inner PNS convection itself may be a source of mode
excitation; see, for example, \citealt{mueller:13}). To clarify the nature of the `gap',
higher resolution simulations are necessary. If this spectrogram feature is real, it
could serve as an interesting analysis tool to probe the structure of the inner PNS.

\subsection{Dependence of the GW signal on parameters}
\label{dependences}

In this section, we outline the key dependences of the GW signal on the
progenitor mass and rotational angular velocity, on the EOS, and on the details of 
the microphysics, such as the inclusion of the many-body corrections to the 
neutrino-nucleon scattering rate and the implementation of the gravity
solver. While qualitatively the GW signal from all our models is very similar,
the frequency of the dominant feature is sensitive to the EOS and the
neutrino-nucleon opacities, and almost insensitive to the progenitor mass.
We apply the analysis of Section~\ref{analysis} to all non-rotating
models from our set, and we confirm the association between the
dominant GW feature and the fundamental $l=2$ mode in each case.

\subsubsection{Dependence of the GW signal on the progenitor mass}
\label{mass}

\begin{figure}
  \centering
  \includegraphics[width=0.45\textwidth]{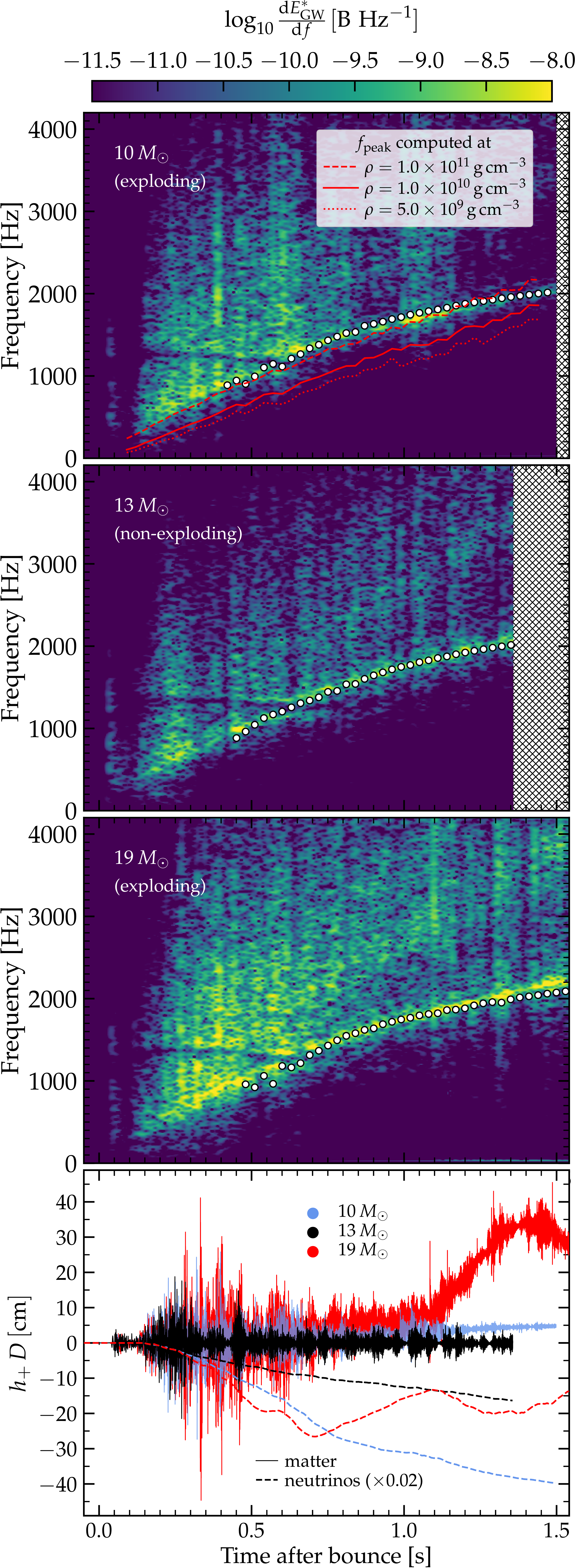}
  \caption{GW spectrograms and waveforms from the models
  \texttt{M10\_SFHo}, \texttt{M13\_SFHo}, and \texttt{M19\_SFHo}, differing
  only in the progenitor mass. White markers show the eigenfrequencies of
  the fundamental quadrupole mode, found as described in Section~\ref{analysis} for each model.
  Gray hatched regions simply fill the blank space left after aligning the simulations in time. 
  Red lines
  in the top panel show the peak GW frequency $f_{\rm peak}$ computed as suggested 
  in \citet{murphy:09} and \citet{mueller:13} (see text for the explanation).} 
  \label{fig:masses}
\end{figure}

Figure~\ref{fig:masses} shows the GW spectrograms and waveforms for
the models \texttt{M10\_SFHo}, \texttt{M13\_SFHo}, and \texttt{M19\_SFHo},
which are simulated with the identical numerical setup and differ only in
the progenitor mass. Two of the models, \texttt{M10\_SFHo} and \texttt{M19\_SFHo}, 
explode at $\sim$400 and $\sim$350$\,{\rm ms}$ after
bounce, respectively, and have a characteristic explosion `tail' in their
waveforms \citep{murphy:09,yakunin:10,mueller:13}. The model \texttt{M13\_SFHo}
does not explode. White markers on the spectrograms
indicate the eigenfrequencies of the fundamental $l=2$ modes, found as 
described in Section~\ref{analysis} for each model. In general, we see good
agreement between the analytical eigenfrequencies and the dominant GW
signal, with the largest deviation seen in the post-explosion phase of the
$19\,M_{\odot}$ model.

For comparison, 
the red lines in the top panel of Figure~\ref{fig:masses} 
show the peak GW frequency $f_{\rm peak}$ computed as suggested in \citet{murphy:09}
and \citet{mueller:13}, where it is associated with the surface value of the Brunt-V\"{a}is\"{a}l\"{a} 
frequency divided by $2\pi$. In Figure~\ref{fig:masses}, we use Equation (17) of \citet{mueller:13}, 
omitting the factor $\left(1-\frac{GM_{\rm PNS}}{R_{\rm PNS}c^2}\right)^2$, where
$M_{\rm PNS}$ and $R_{\rm PNS}$ are the mass and radius of the PNS, 
respectively. As in \citet{pan:17}, we find that removing this factor results in better
agreement between $f_{\rm peak}$ and the dominant feature of the GW
spectrogram\footnote{This may be related to the fact
that we use the approximate relativistic gravitational potential in
our simulations.}. The three lines correspond to the three different density
isosurfaces of $\rho = 5.0\times10^{9}\,{\rm g}\,{\rm cm}^{-3}$,
$10^{10}\,{\rm g}\,{\rm cm}^{-3}$, and $10^{11}\,{\rm g}\,{\rm cm}^{-3}$, which 
can represent the PNS surface.
The plot shows that using the conventional definition for the PNS
surface, $\rho = 10^{11}\,{\rm g}\,{\rm cm}^{-3}$, the analytical formula
for $f_{\rm peak}$ may provide a good fit to the dominant GW signal.
At the same time, the value of $f_{\rm peak}$ is sensitive to the location of 
the PNS surface, which currently lacks strict physical definition.

\begin{figure}
  \centering
  \includegraphics[width=0.48\textwidth]{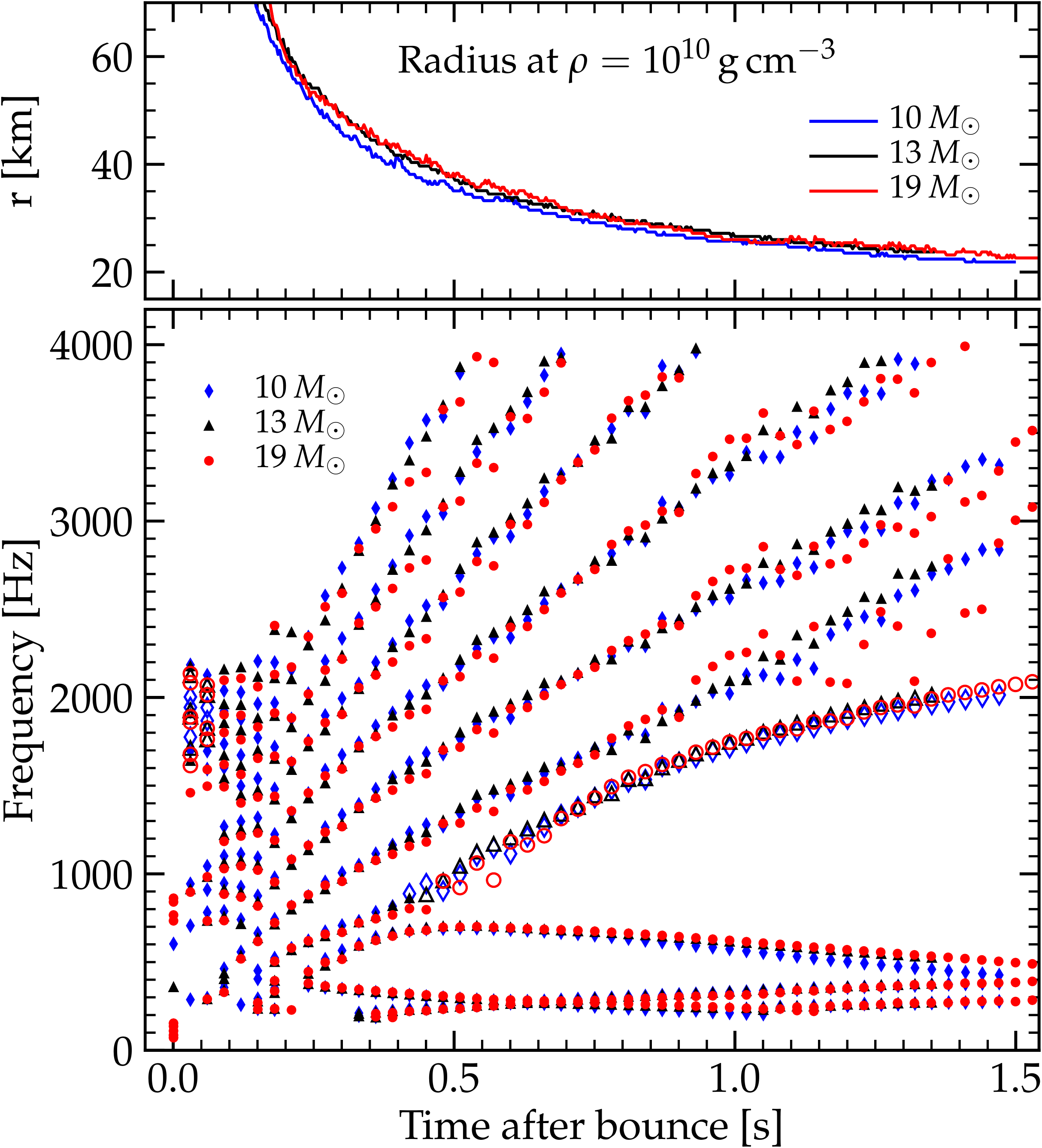}
  \caption{Top panel: The radius at which the angle-averaged density of the models
  \texttt{M10\_SFHo}, \texttt{M13\_SFHo}, and \texttt{M19\_SFHo}
  is equal to $10^{10}\,{\rm g}\,{\rm cm}^{-3}$. This represents the outer boundary
  in the linear perturbation analysis of Section~\ref{analysis}, and it can be used as a
  proxy for the PNS radius (although $\rho=10^{11}\,{\rm g}\,{\rm cm}^{-3}$ is more
  commonly used in the literature for that). Bottom panel:
  $l=2$ eigenfrequencies of these models, calculated using linear perturbation analysis, as described in
  Section~\ref{analysis}. Large empty symbols represent the fundamental (0-nodes) mode,
  which is also shown in Figure~\ref{fig:masses}. This plot demonstrates that the
  dominant frequency of the GW signal depends weakly on the progenitor ZAMS mass.} 
  \label{fig:l2masses}
\end{figure}

The agreement between the analytic eigenfequencies and the GW 
spectrograms allows us to compare the spectrograms by comparing the
frequencies. Figure~\ref{fig:l2masses} shows the full results of the linear perturbation
analysis for the three considered models, performed as in 
Section~\ref{analysis}. In the bottom panel of Figure~\ref{fig:l2masses}, 
the filled symbols show the modes
with larger than zero number of nodes, while the empty symbols show the
f-mode frequencies. This plot demonstrates that the PNS
eigenfrequencies in general, and the frequencies of the fundamental 
quadrupolar mode in particular,
are strikingly similar between the models, despite the large difference in their
progenitor masses and even in the waveforms themselves. 

This may reflect the fact that the evolution of the PNS radius is very similar
between the models with different progenitor masses, which is shown in the top
panel of Figure~\ref{fig:l2masses}, and was already noticed in the literature
for a wide range of progenitors differing not only by the ZAMS mass, but also in the
metallicity (see, for example, Figure 7 of \citealt{bruenn:16}, Figure 10 of 
\citealt{summa:16}, Figure 15 of \citealt{radice:17})\footnote{Note that in the
top panel of Figure~\ref{fig:l2masses} we show the radii where the angle-averaged
density is equal to $10^{10}\,{\rm g}\,{\rm cm}^{-3}$, which also serves as the
outer boundary in our analysis. It is more common in the literature to use 
$\rho=10^{11}\,{\rm g}\,{\rm cm}^{-3}$ as the definition of the PNS radius. 
In our models, the radii at the density $10^{11}\,{\rm g}\,{\rm cm}^{-3}$ are
nearly the same as the radii at the density $10^{10}\,{\rm g}\,{\rm cm}^{-3}$.}.
Indeed, if the GW signal from CCSNe is so tightly related to the
PNS eigenmodes, the structure of the PNS should be the main factor
defining the time-frequency structure of this signal.

\begin{figure}
  \centering
  \includegraphics[width=0.475\textwidth]{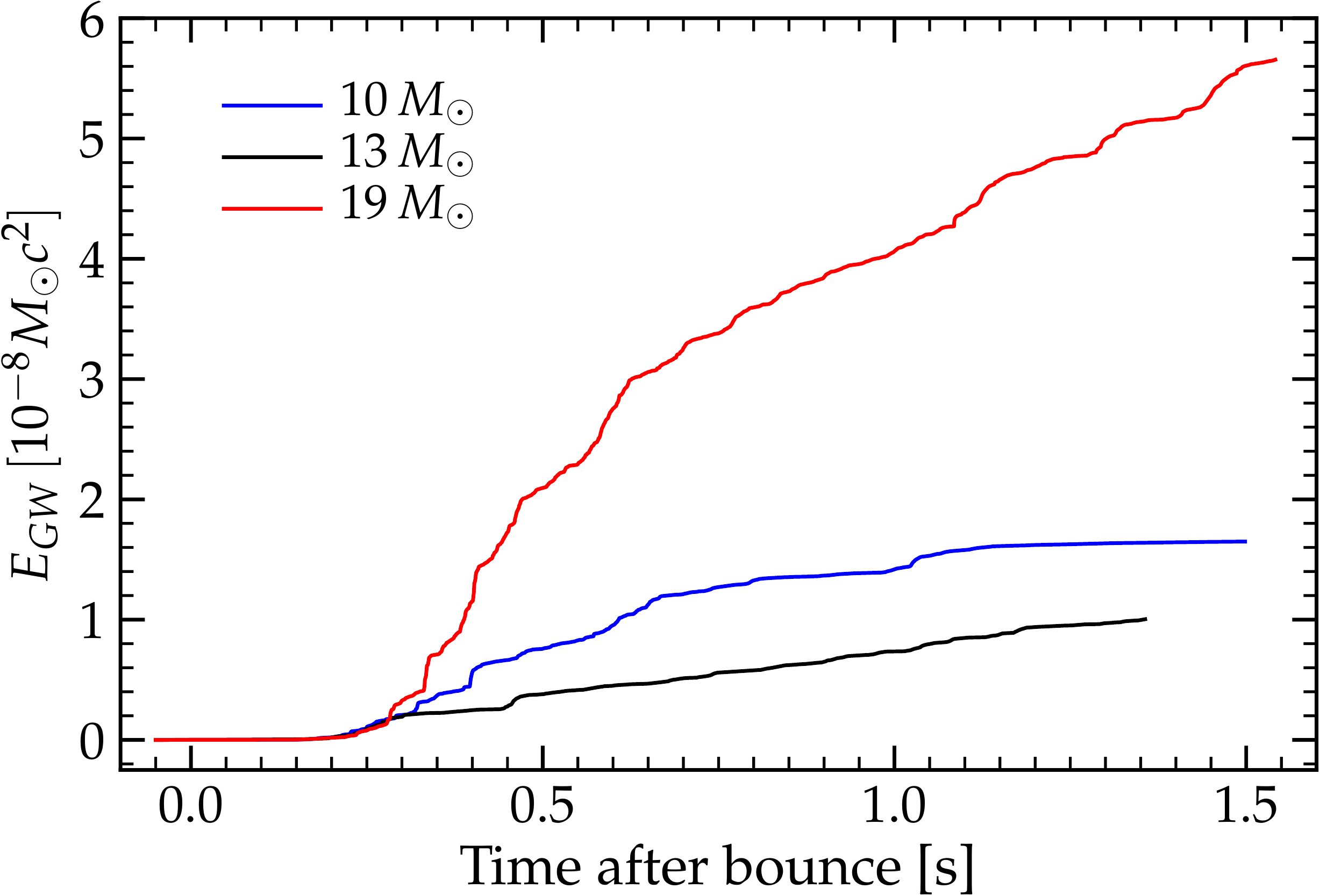}
  \caption{Total energy emitted in GWs from models 
  \texttt{M10\_SFHo}, \texttt{M13\_SFHo}, and \texttt{M19\_SFHo}
  as a function of time. The dependence of $E_{\rm GW}$ on the progenitor
  ZAMS mass is not monotonic.} 
  \label{fig:egw}
\end{figure}

The same argument cannot be applied to the amplitude of the GW signal, 
which, instead, must depend on the mechanism of excitation of the PNS
modes. It was shown in many previous studies that the GW signal from
CCSNe experiences sudden increases in amplitude at the moments
when the PNS surface is hit by the downfalling accretion `plumes' 
\citep{murphy:09,mueller:13,yakunin:15}. It is, therefore, natural to expect
that the GW power will depend on the details of the postshock accretion, 
which takes place above the PNS surface and is largely
determined by the core structure of the progenitor. Figure~\ref{fig:egw}
shows the energy emitted in GWs due to the matter motions alone as a
function of time for the models 
\texttt{M10\_SFHo}, \texttt{M13\_SFHo} and \texttt{M19\_SFHo}. In these
models, we don't see a monotonic dependence of the GW power 
on the progenitor mass, with the model \texttt{M13\_SFHo} 
producing the weakest signal among the three. 
In fact, it is hard to expect such a monotonic dependence, 
because the dependence of the progenitor core structure itself 
on the progenitor ZAMS mass is not monotonic \citep{sukhbold:16} and, 
moreover, may be intrinsically chaotic \citep{sukhbold:17}.
For this reason, we advise using caution when deducing
the dominant signal frequency based on the total GW energy spectrum, 
especially if it is done for the purpose of comparing models with different progenitor
masses. Accretion downflows hitting the PNS surface at random moments
of time may give more weight to the system eigenfrequencies in those
moments, complicating the overall picture. Instead, the comparison of the 
time-frequency spectrograms serves this purpose best.

\begin{figure}
  \centering
  \includegraphics[width=0.49\textwidth]{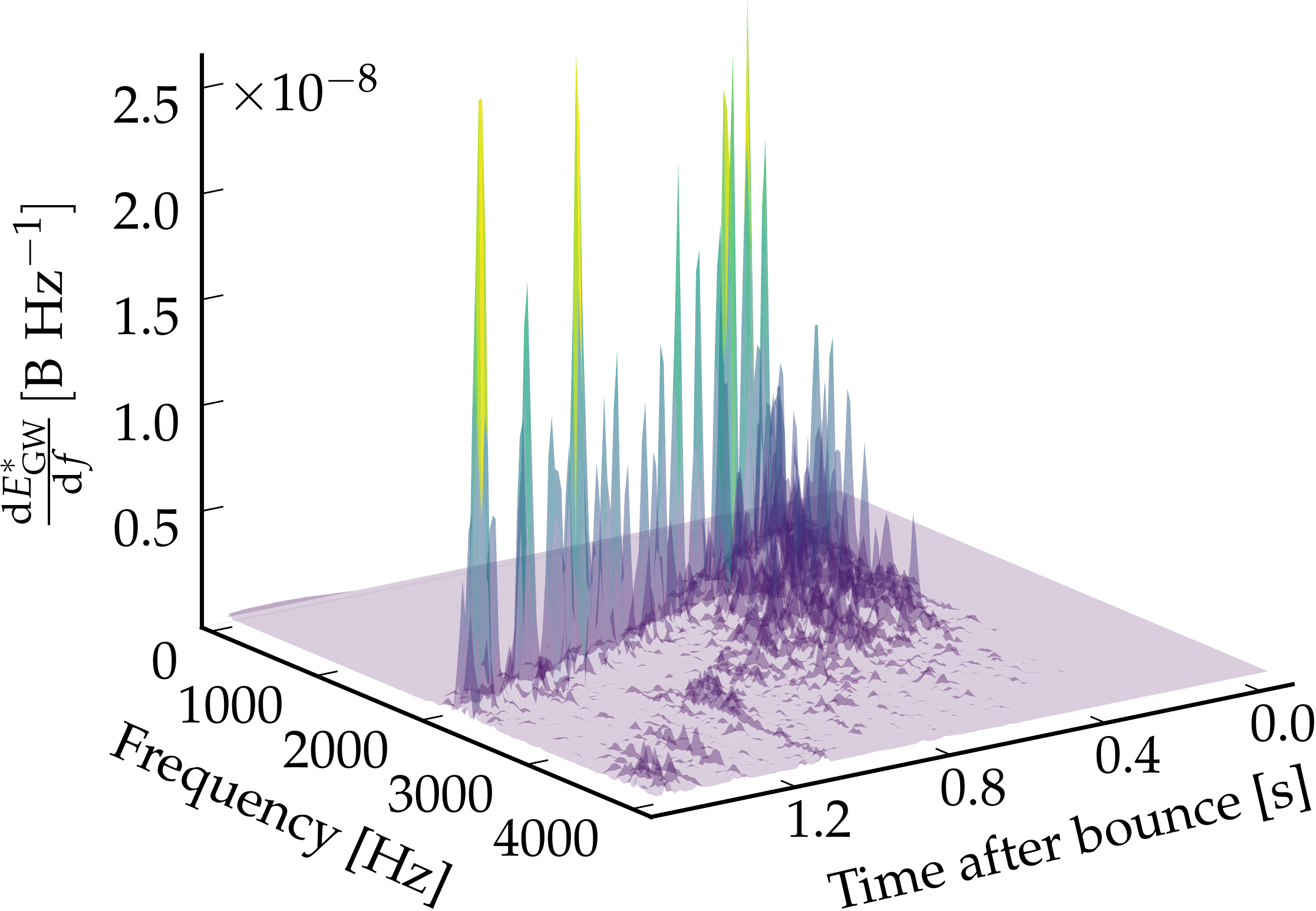}
  \caption{Linear 3D representation of the GW spectrogram from the 
  model \texttt{M19\_SFHo}. This model starts exploding at $\sim$350$\,{\rm ms}$
  after the core bounce, but the dominant component of the GW signal does not
  decay and stays strong until the end of the simulation, for more than a second after
  the explosion.} 
  \label{fig:3d19}
\end{figure}

The strongest signal among all our models is produced by the model 
\texttt{M19\_SFHo}. In Figure~\ref{fig:3d19}, we present the linear 3D 
representation of the GW spectrogram from this model. This figure emphasizes the
point made in Section~\ref{example}, that the GW signal may stay
strong for a long time after the explosion (more than a second in the case
of \texttt{M19\_SFHo}). The large offset from zero seen
in the GW strain of this model at late times (the bottom panel of Figure~\ref{fig:masses})
suggests a very asymmetric character for its explosion. This is indeed the case,
as demonstrated in Figure 6 of \citet{vartanyan:18}, which shows 
snapshots of the electron fraction
and entropy of this model at different moments of time. For the analogous snapshots
of the $10\,M_{\odot}$ model, we refer the reader to \citet{radice:17}.

The dependence of the CCSN GW signal on the progenitor mass was 
previously studied in a number of works 
\citep{murphy:09,mueller:13,yakunin:15}.
For example,
\citet{mueller:13} report $\sim$30$\%$ differences in the typical emission frequencies
between their $11.2$ and $25\,M_{\odot}$ models, which they admit to be
small for this large a mass difference. Our comparison, however, shows even
smaller scatter, no more than $\sim$5$-$10$\%$ in frequency across the considered mass range,
without any systematic trend. On one hand, we cannot
exclude that at least part the difference between the $11.2$ and $25\,M_{\odot}$ 
models of \citet{mueller:13} may come from the fact that they were simulated with a
slightly different EOS (we discuss the dependence of the signal on EOS in the
next subsection). On the other hand, the models of \citet{mueller:13} 
treat general relativity more accurately by solving the relativistic equations of
hydrodynamics in the conformally-flat approximation, 
while our work uses the effective potential approach, which
may also affect the dominant frequency of the signal 
\citep[see][]{mueller:13}.

\subsubsection{Dependence of the GW signal on the equation of state}
\label{eos}

\begin{figure}
  \centering
  \includegraphics[width=0.46\textwidth]{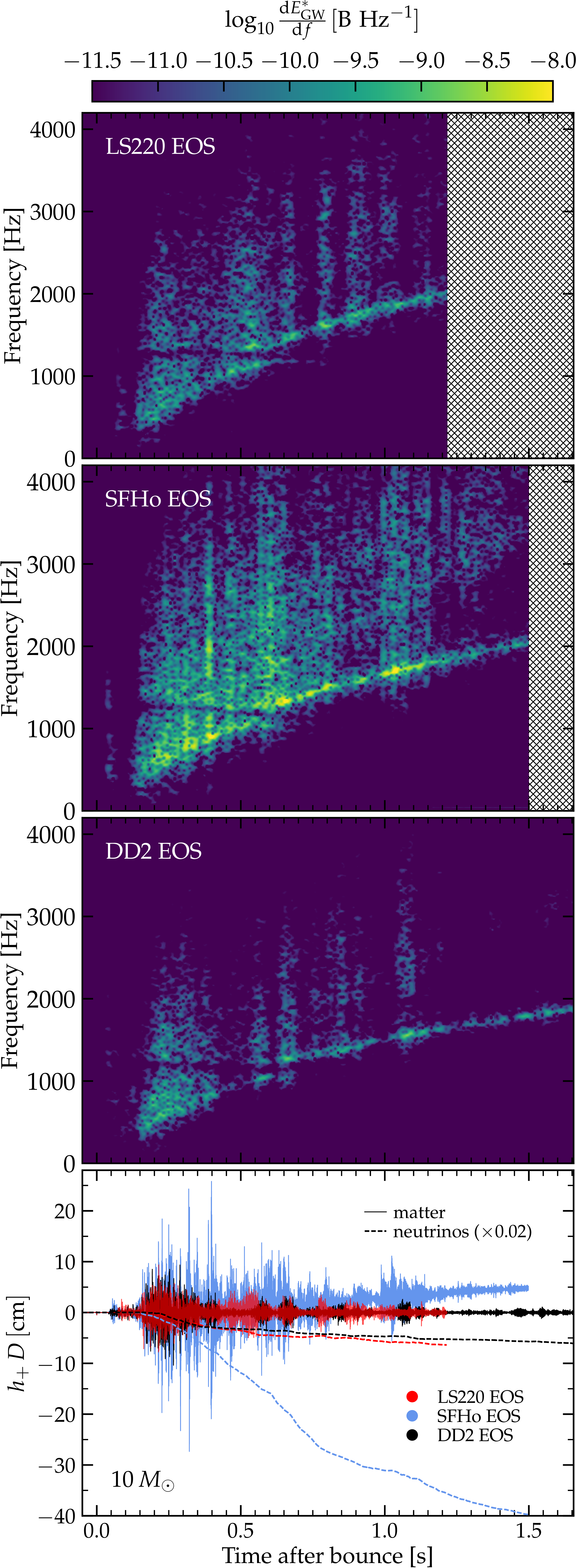}
  \caption{GW spectrograms and waveforms from the models \texttt{M10\_LS220},
   \texttt{M10\_SFHo}, and  \texttt{M10\_DD2}, differing
  only in the EOS. Gray hatched regions simply fill the blank space 
  left after aligning the simulations in time.} 
  \label{fig:eos}
\end{figure}

\begin{figure}
  \centering
  \includegraphics[width=0.465\textwidth]{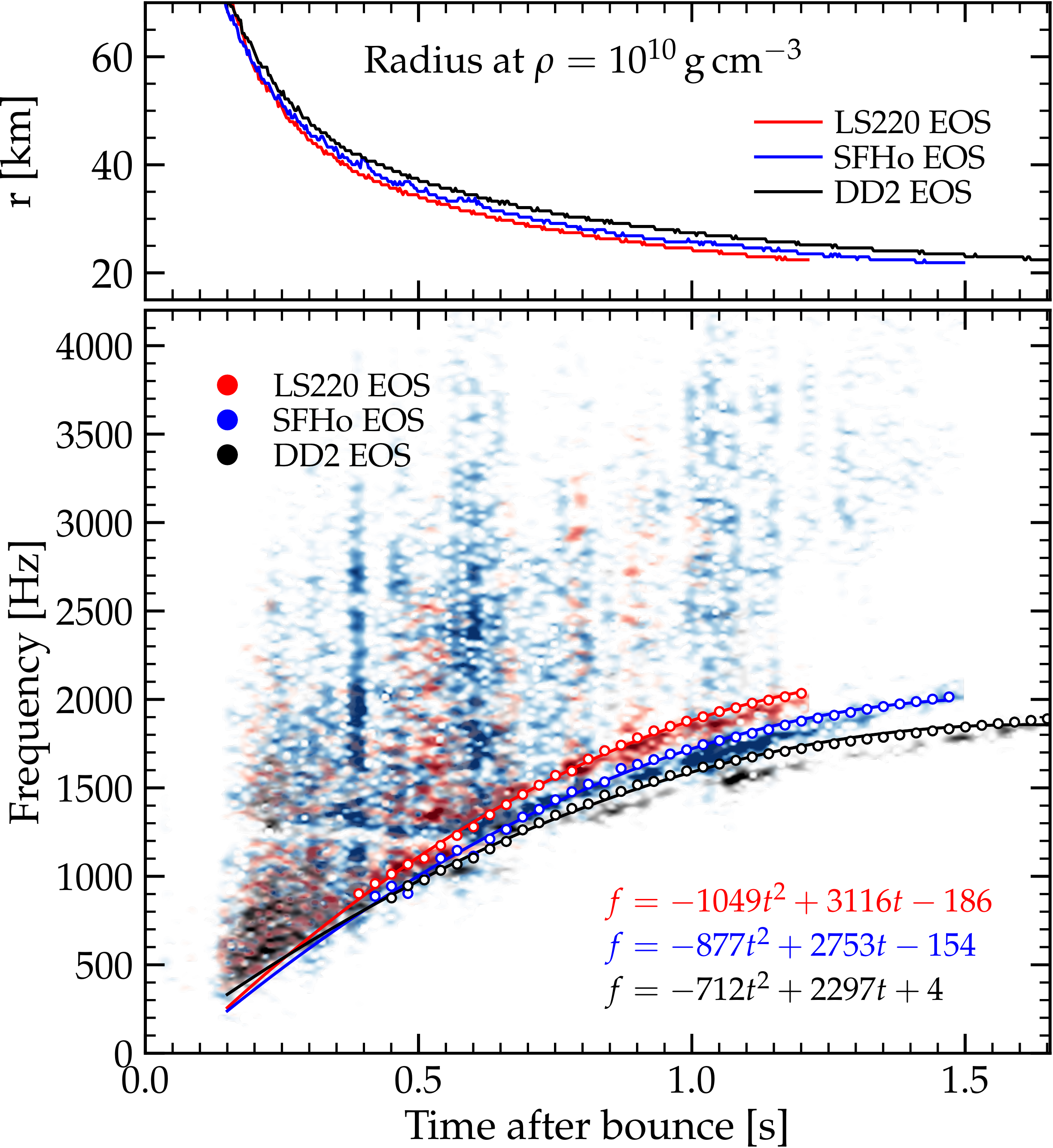}
  \caption{Top panel: The radius at which the angle-averaged density of the models
  \texttt{M10\_LS220}, \texttt{M10\_SFHo}, and \texttt{M10\_DD2}
  is equal to $10^{10}\,{\rm g}\,{\rm cm}^{-3}$. This represents the outer boundary
  in the linear perturbation analysis of Section~\ref{analysis}, and it can be used as a
  proxy for the PNS radius (although $\rho=10^{11}\,{\rm g}\,{\rm cm}^{-3}$ is more
  commonly used in the literature for that). 
  Bottom panel: Comparison of the GW spectrograms from these models, differing
  only in the EOS. Empty markers of the corresponding color show
  the f-mode eigenfrequencies and
  demonstrate that the linear perturbation analysis captures well the
  dependence of the dominant feature of the GW spectrogram on the EOS. The
  lines represent the second-order polynomial fits of the f-mode eigenfrequencies,
  and the explicit form of the fits is given in the right bottom corner (there, $f$ 
  is the frequency in Hz and $t$ is the time in seconds). These can
  be used as a prior in the search of CCSN GW signal with the ground-based laser
  interferometers.} 
  \label{fig:eos_analysis}
\end{figure}

Figure~\ref{fig:eos} shows the GW spectrograms and waveforms for
the models \texttt{M10\_LS220}, \texttt{M10\_SFHo}, and \texttt{M10\_DD2}, 
which were
simulated with three different EOSes. All other numerical parameters and
the details of microphysics are the same between these models. We find
that the EOS has a large impact on the amplitude of the GW signal, 
its dominant frequency, total emitted
energy $E_{\rm GW}$ (see Table~\ref{tab:models}), and even the qualitative outcome of 
the simulation (the model \texttt{M10\_SFHo} explodes, unlike the other two). 
Similar EOS sensitivity of the simulation outcome was recently reported
by \citet{pan:17} in the context of the GW signal from black hole formation in failed SNe.

In order to emphasize the dependence of the dominant GW frequency on the
EOS, we overplot the GW spectrograms of these models in the bottom
panel of Figure~\ref{fig:eos_analysis}. Empty markers represent the f-mode
eigenfrequencies found from the linear perturbation analysis of the
models, as described in Section~\ref{analysis}. The top panel
of Figure~\ref{fig:eos_analysis} shows the evolution of the PNS radii taken at
the value of density $10^{10}\,{\rm g}\,{\rm cm}^{-3}$. Compared to the top
panel of Figure~\ref{fig:l2masses}, the difference between the 
PNS radii in Figure~\ref{fig:eos_analysis} is slightly
larger and more systematic, which translates into the systematic $\sim$10$-$15$\%$
difference in the dominant frequencies of the GW signal, which, in turn, is
well captured by our analysis (the largest disagreement is seen in the 
\texttt{M10\_DD2} model).
Interestingly, among the three EOSes used in our study, SFHo is
the `softest' one, while DD2 is the `hardest'. Nevertheless, the smallest PNS
radius and the largest GW frequency are produced by the LS220 EOS. 
This suggests that the EOS dependence of the GW signal, as well as the overall core
evolution, may not necessarily be described in terms of a single stiffness parameter
defined at zero temperature.

\begin{figure}
  \centering
  \includegraphics[width=0.465\textwidth]{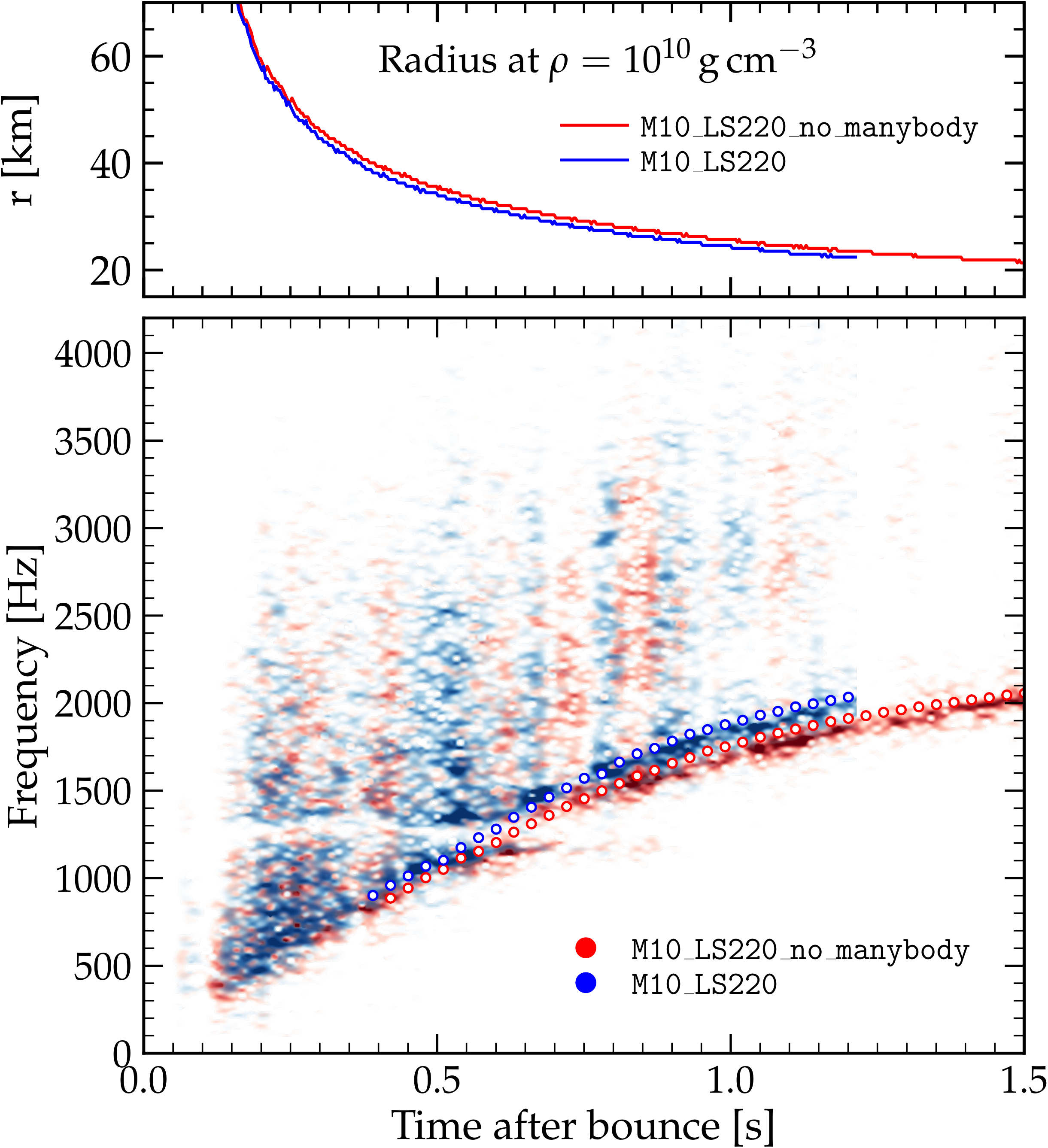}
  \caption{Comparison of the GW spectrograms from the models simulated
  with (blue, \texttt{M10\_LS220}) and without 
  (red, \texttt{M10\_LS220\_no\_manybody}) the many-body corrections to
  the neutrino-nucleon scattering rates. Empty markers of the corresponding color show
  the f-mode eigenfrequencies.} 
  \label{fig:horo}
\end{figure}

We quantify the dependence of the f-mode eigenfrequency on time for models
\texttt{M10\_LS220}, \texttt{M10\_SFHo}, and \texttt{M10\_DD2} by fitting
it with a polynomial. We find that a simple quadratic function in the form
$f=At^2+Bt+C$, where $f$ is frequency in Hz, $t$ is time in seconds, and $A$, $B$
and $C$ are coefficients, adequately describes the dependence
over the first $\gtrsim$1.5 seconds after bounce, while the core keeps
shrinking. Eventually, the PNS will cool down and deleptonize, which could lead to the
flattening of the frequency-time curve.
The quadratic fits are shown with the lines of corresponding color in Figure~\ref{fig:eos_analysis}
and explicitly written down in the right bottom corner of the figure. 
These fits can be used as priors when looking
for the CCSN GW signal in the data from ground-based laser interferometers, such
as LIGO, Virgo, or KAGRA. At the same time, we emphasize that the accuracy of these fits 
may be affected by the details of the physics and microphysics used in our (and other) codes. 
For example, to demonstrate the
sensitivity of the GW signal to the details of the neutrino opacity, we compare the
spectrograms from models \texttt{M10\_LS220} and \texttt{M10\_LS220\_no\_manybody} 
in Figure~\ref{fig:horo}. The many-body corrections
to the neutrino-nuclear scattering cross section decrease the neutrino opacity, which
leads to the faster contraction of the PNS, as shown in the top panel of Figure~\ref{fig:horo}.
The bottom panel of Figure~\ref{fig:horo} shows that neglecting 
these corrections results in a $\sim$10$\%$ shift in the 
dominant GW frequency. Another factor influencing the GW frequency is the description
of the gravitational field \citep[see, e.g.][]{mueller:13}. Taking all these factors into account,
we expect the accuracy of the fits from Figure~\ref{fig:eos_analysis} to be not worse than
$\sim$30$\%$.

On the other hand, the power of the GW signal does demonstrate monotonic dependence
on the stiffness of the EOS, with the hardest EOS (DD2 in our case) 
producing the weakest signal. As we already mentioned in the previous subsection, 
the amplitude of the GW signal is largely determined by accretion and 
post-shock convection, which act as driving forces for the excitation of the PNS
oscillations. It was found in previous work 
\citep{marek:09,kuroda:16,kuroda:17} that softer EOSes
result in more vigorous SASI activity. While we do not clearly identify SASI in any of
the three models, we also find that the shock oscillations are strongest in the
\texttt{M10\_SFHo} model and weakest in the \texttt{M10\_DD2} model. This leads
to stronger excitation of the PNS modes and more powerful GW signals in case of
the softest EOS.

\subsubsection{Dependence of the GW signal on rotation}
\label{rotation}

\begin{figure}
  \centering
  \includegraphics[width=0.41\textwidth]{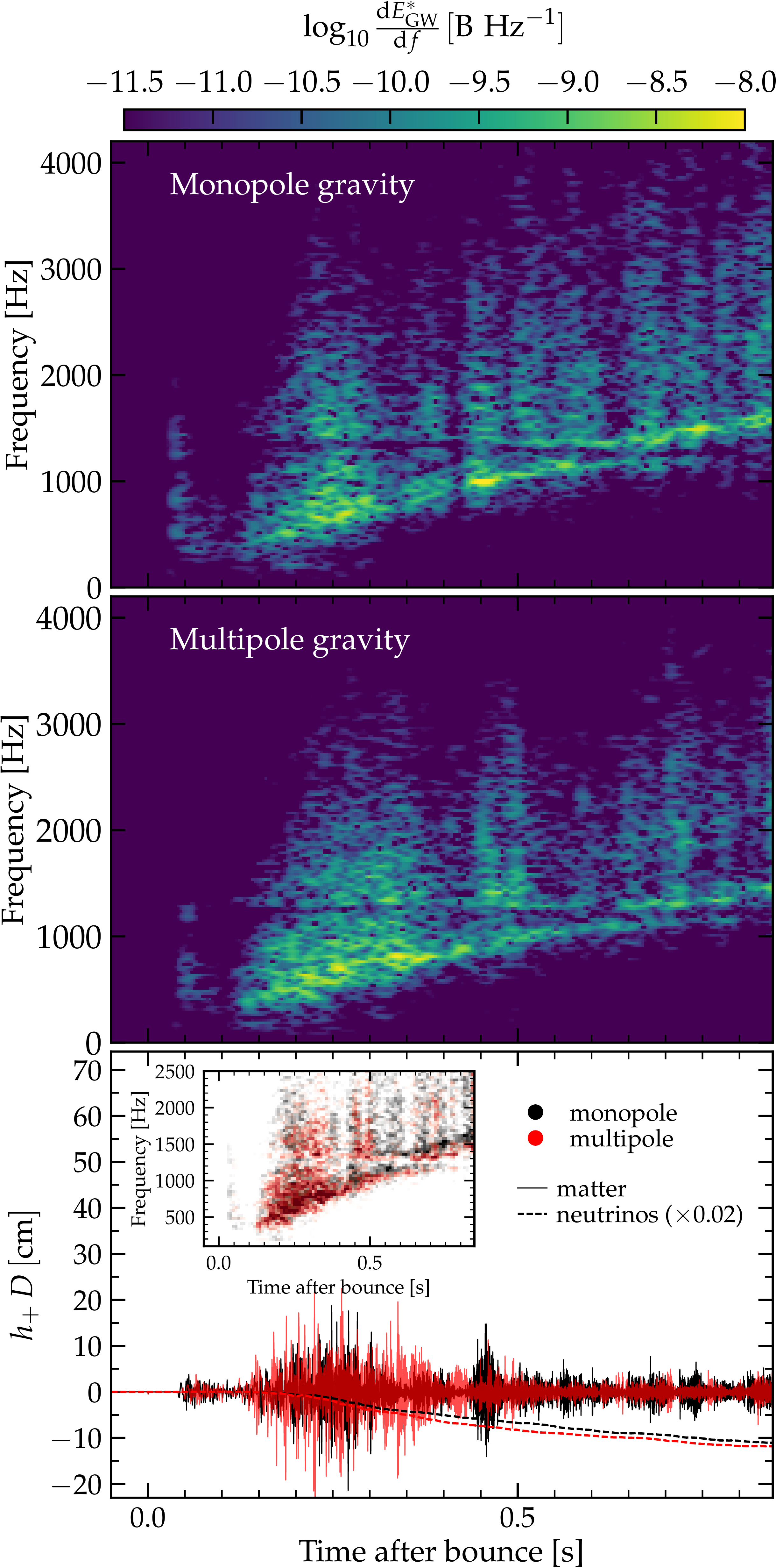}
  \caption{GW spectrograms and waveforms of the models \texttt{M13\_SFHo} 
  and \texttt{M13\_SFHo\_multipole}, differing only in the gravity implementation
  (see Section~\ref{setup}).} 
  \label{fig:multi}
\end{figure}

Simulations of rotating core collapse were the first to predict and study
the GW emission from CCSNe 
\citep{ott:07,ott:12,dimmelmeier:07,dimmelmeier:08,abdikamalov:10}. Because of
the symmetry breaking introduced by rotation, these models produce strong
GW signals already at the early stages of collapse and bounce, which makes
even short (few tens of milliseconds) simulations very informative.
Not very demanding in terms of the neutrino physics, these simulations 
progressed enough to establish the connection between the properties of the
GW signal and the progenitor core parameters 
\citep{summerscales:08,logue:12,roever:09,abdikamalov:14,engels:14,fuller:15,powell:16,richers:17}.
The main limitation of these papers is that fast rotating cores are
not very common among CCSN progenitors \citep{heger:05,woosley:06}.
Here, we focus on the GW signal from a moderately 
($\Omega_0=0.2\,{\rm rad}\,{\rm s}^{-1}$) rotating progenitor, and
follow it for a full second after bounce, which, to the best of our knowledge, is
currently the longest simulation of its kind, for which the GW signal has been extracted.


In the rotating model, we use the multipole gravity solver of \citet{mueller:95}.
For all other models shown before, we used a monopole approximation
for the gravitational potential \citep{marek:06}. As an aside, to show how the gravity implementation
alone influences the GW signal, we compare the spectrograms and waveforms from the
models \texttt{M13\_SFHo} and \texttt{M13\_SFHo\_multipole} in Figure~\ref{fig:multi}.
The difference between the models is noticeable, although not large, resulting in
$\sim$10$\%$ shift in the dominant frequency by the end of the \texttt{M13\_SFHo\_multipole} 
simulation. This tells us that the full general-relativistic approach to gravity (which is, strictly speaking,
the only correct approach) is important for the accurate quantitative description of the GW signal.

\begin{figure}
  \centering
  \includegraphics[width=0.41\textwidth]{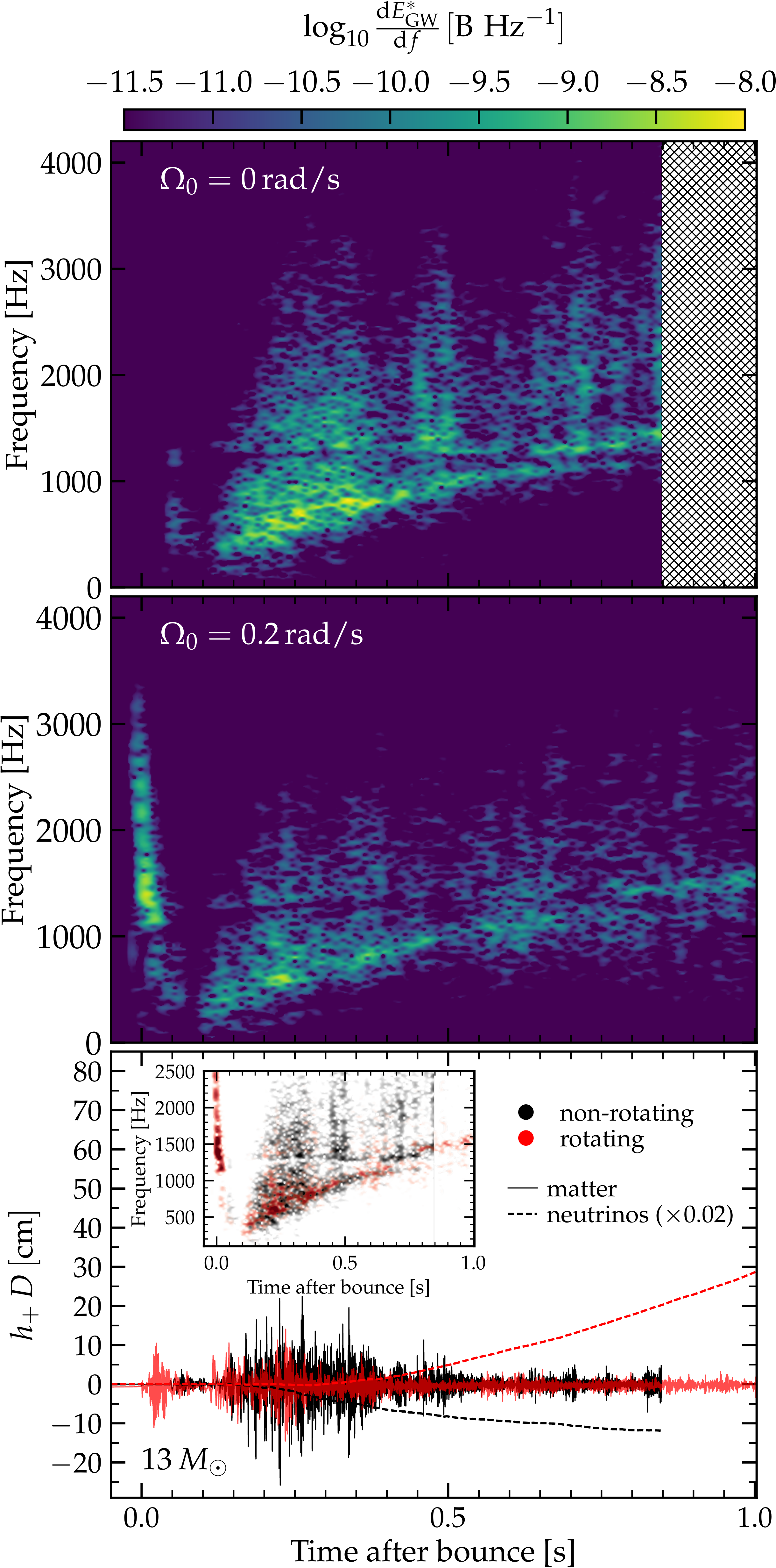}
  \caption{Comparison of the GW spectrograms and waveforms from the models
  \texttt{M10\_SFHo\_multipole} and \texttt{M10\_SFHo\_rotating}, differing only in
  the angular velocity. Gray hatched regions simply fill the blank space 
  left after aligning the simulations in time.} 
  \label{fig:rot}
\end{figure}

Figure~\ref{fig:rot} shows the GW spectrograms and the waveforms from
the non-rotating (\texttt{M10\_SFHo\_multipole}) and rotating (\texttt{M10\_SFHo\_rotating})
models, which have identical numerical setups, apart from the angular velocity.
In agreement with the previous literature, the rotating model generates a strong
GW signal at the core bounce, lasting for a few tens of milliseconds, followed by
the short quiescent phase. At the same time, the main component of the GW signal
is noticeably weaker for this model, though the dominant frequency does not
seem to change much.
Interestingly, the `gap' still persists in the GW spectrograms of both models.


\section{Conclusions and discussion}
\label{discussion}

The main findings of our study can be briefly summarized as follows:
\begin{itemize}
\item We reproduce the dominant, long-lasting GW signal from CCSNe
by means of linear perturbation analysis and associate it with a g-mode having
2 radial nodes at the early stage ($\sim$200$-$400$\,{\rm ms}$ after bounce)
and with the f-mode later on (from $\sim$400$\,{\rm ms}$ until more than a 
second after bounce). This finding presages future opportunities for
the analytical study of the CCSN GW signal.

\item We demonstrate a weak dependence of the dominant GW frequency
on the progenitor ZAMS mass and provide a simple quadratic fit for it as a 
function of time for three different EOSes. This may help identify possible
CCSN candidates in the GW data from ground-based laser interferometers.

\item We identify a new feature in the GW spectrogram, which looks like a
`gap' across the noisy GW emission in the first $\sim$200$-$700$\,{\rm ms}$ 
after bounce. Our attempts to explain it as a numerical artifact failed.
We explain the `gap' as the interaction between the outer p-modes
and g-modes of the PNS inner ($\sim$10$\,{\rm km}$) core, probably as 
a result of avoided crossing.

\item We show the effect of moderate ($0.2\,{\rm rad}\,{\rm s}^{-1}$) initial 
progenitor rotation on the GW signal. The rotation strengthens the
bounce signal, but weakens the dominant part of the post-bounce GW emission.
\end{itemize}

All simulations analyzed in our study are 2D, which raises a question
how our conclusions will change in the full 3D case. It is known from
previous studies that the success of an explosion in the CCSN simulations
largely depends on the hydrodynamical instabilities and the associated
turbulent pressure behind the stalled shock 
\citep{burrows:95,murphy:13,couch:15,mueller:15,abdikamalov:16,takahashi:16,mueller:17},
which also increases the exposure of matter to neutrino heating
\citep{buras:06b,murphy:08}. However,
it is known that turbulence has different properties in 2D and
3D \citep{kraichnan:67}, and it has been shown that this difference
artificially facilitates explosion \citep{hanke:12,dolence:13,takiwaki:14,
couch:14,couch:15,abdikamalov:15}. Therefore, if the properties of
turbulence in the gain region were directly reflected in the GW
spectrogram, we would expect it to differ in 3D. 
Instead, our analysis suggests that the strongest 
component of the GW signal is associated with the fundamental mode of
the PNS itself, which is expected to be nearly spherical even in the
3D case. Turbulence in this case acts only as a driving force exciting the mode
oscillations. This makes us believe that the time-frequency structure
of the GW signal shown here and its linear analysis will still be applicable
for 3D models, while the amplitude may change (become smaller). 
The same was recently suggested in \citet{yakunin:17}, 
where the authors obtained similar behavior of the GW signal for a 2D and a 3D model.
In their 3D case, convection was characterized by a larger number of
relatively small scale structures, as opposed to the few massive accretion funnels in 
2D. This led in 3D to smoother GW energy emission, but caused only moderate 
changes in its spectral distribution, vis-\`{a}-vis their 2D results, during 
the first $450\,{\rm ms}$ of the signal. More about the comparison between
the 2D and 3D GW signals from CCSNe may be found in \citet{andresen:17}.

Interestingly, our linear analysis presents an opportunity to predict the dominant
frequency of the GWs from CCSNe based on 1D simulations.
However, this approach should be applied
with great caution, because, for example, the evolution of the PNS radius 
differs between the 1D and 2D simulations for the same models 
\citep{radice:17}. At the same time, such an analysis allows one to quickly
cover large regions of parameter space related to the EOS and microphysics, in order
to investigate which of the parameters has the strongest influence on the GW signal.

\citet{pan:17} suggested that increasing the sensitivity of the
next generation GW detectors in the $\sim$1000$\,{\rm Hz}$ window is very important
for the study of the BH formation in failed SNe 
\citep[see also][]{kuroda:18}. We add to this statement that
high frequency sensitivity is crucial for the detection of the GW signal
from the successful SN explosions as well. Increasing the sensitivity of
aLIGO and KAGRA in this band would help us to fully exploit the luck
of the next nearby SN discovery and trace the high-frequency GW signal of a
newborn NS.

\acknowledgments
We thank Aaron Skinner and James Stone for helpful discussions and feedback. We
thank Pablo Cedr\'{a}-Dur\'{a}n and Jos\'{e} Antonio Font for finding an error in the original
calculations, and for other helpful suggestions.
The authors would like to acknowledge support of the U.S. NSF under award AST-1714267, the Max-Planck/Princeton Center (MPPC) for Plasma Physics (under award NSF PHY-1144374), and the DOE SciDAC4 Grant DE-SC0018297 (under subaward 00009650). The authors employed computational resources provided by the TIGRESS high performance computer center at Princeton University, which is jointly supported by the Princeton Institute for Computational Science and Engineering (PICSciE) and the Princeton University Office of Information Technology. They also acknowledge a supercomputer allocation by the National Energy Research Scientific Computing Center (NERSC), which is supported by the Office of Science of the US Department of Energy (DOE) under contract DE-AC03-76SF00098. DR acknowledges support from a Frank and Peggy Taplin Membership at the Institute for Advanced Study and the Max-Planck/Princeton Center (MPPC) for Plasma Physics (NSF PHY-1523261).

\bibliographystyle{apj}


\renewcommand \theequation {A.\arabic {equation}}
\setcounter{equation}{0}

\appendix

\section{Resolution dependence of the GW signal for the model \texttt{M10\_LS220\_no\_manybody}.}
\label{Ap0}

\begin{figure*}[!htb]
  \centering
  \includegraphics[width=0.99\textwidth]{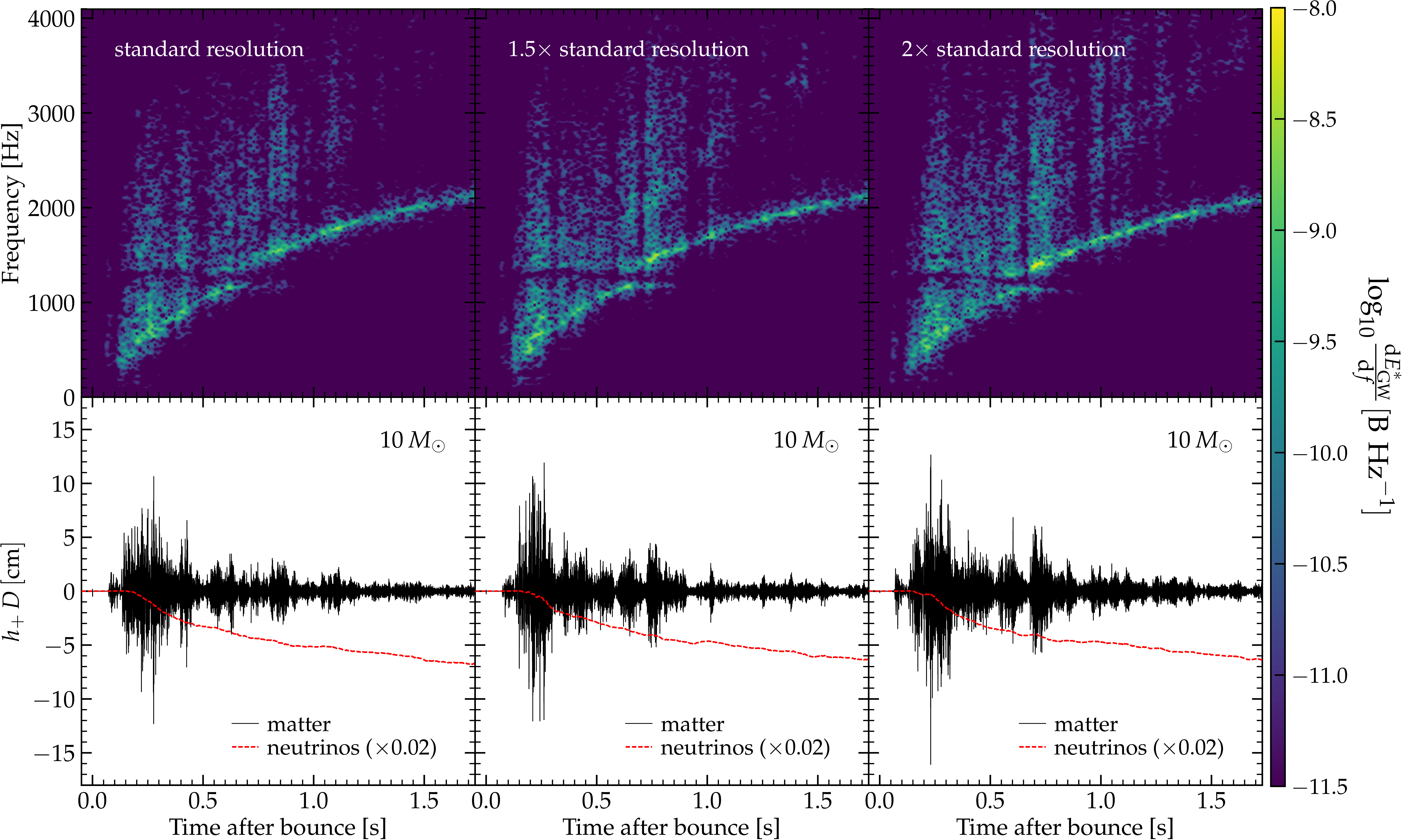}
  \caption{Spectrograms (top) and the corresponding waveforms (bottom) of 
  the GW signal for the model \texttt{M10\_LS220\_no\_manybody}
  (non-exploding) model for three different levels of resolution.} 
  \label{fig:resolution}
\end{figure*}

Figure~\ref{fig:resolution} shows the GW spectrograms and waveforms
of the model \texttt{M10\_LS220\_no\_manybody} for three different grid resolutions, of which the
lowest (`standard') is used in all other models of this study. It is clear from the figure
that the overall structure of the GW signal and its spectrogram depend
weakly on resolution, demonstrating the robustness of our
results. 

\section{Derivation of the linear perturbation equations including
lapse variation}
\label{Ap1}

Here, we derive the set of equations describing linear perturbations
of a spherically-symmetric background, including the perturbation of the lapse
function. Generally, our calculations follow the same scheme as described
in \citet{torres:17}, and for simplicity of comparison we use the same
notation where possible. All equations are given in geometrized units. 
We start with a static spherically-symmetric 
conformally flat space-time metric in isotropic coordinates $(t,x^i)$:
\begin{equation}
ds^2=g_{\mu \nu}dx^{\mu}dx^{\nu} = - \alpha^2 dt^2 + \psi^4 f_{i j}dx^i dx^j\ ,
\end{equation}
where $\alpha$ is the lapse function, $\psi$ is the conformal factor, and 
$f_{i j}$ is the flat spatial 3-metric. In this metric, the equations of 
general-relativistic hydrodynamics for a perfect fluid can be rendered in the form 
\citep{torres:17,banyuls:97}:
\begin{eqnarray}
&&\label{hydro1}\frac{1}{\sqrt{\gamma}}\partial_t \left[\sqrt{\gamma}D\right]
+ \frac{1}{\sqrt{\gamma}} \partial_i \left[\sqrt{\gamma}D\nu^{*i}\right] = 0\ , \\
&&\label{hydro2}\frac{1}{\sqrt{\gamma}}\partial_t \left[\sqrt{\gamma}S_j\right]
+ \frac{1}{\sqrt{\gamma}} \partial_i \left[\sqrt{\gamma}S_j\nu^{*i}\right] 
+ \alpha\partial_i P = \frac{\alpha \rho h}{2} u^{\mu} u^{\nu} 
\partial_j g_{\mu\nu}\ , \\
&&\label{hydro3}\frac{1}{\sqrt{\gamma}}\partial_t \left[\sqrt{\gamma}E\right]
+ \frac{1}{\sqrt{\gamma}} \nabla_i \left[\sqrt{\gamma}\left( E\nu^{*i}
+ \alpha P \nu^i \right)\right] = \alpha^2 \left(T^{\mu 0} \partial_{\mu} \ln\alpha 
- T^{\mu\nu} \Gamma^{0}_{\mu\nu} \right)\ .
\end{eqnarray}
Here, $T^{\mu\nu} = \rho h u^{\mu} u^{\nu} + P g^{\mu\nu}$ is the 
energy-momentum tensor of a perfect fluid, where $\rho$ is its rest-mass
density, $P$ is the pressure, $u^{\mu}$ is the 4-velocity, 
$h \equiv 1 + \epsilon + P/\rho$ is the specific enthalpy, and $\epsilon$
is the specific internal energy. $\Gamma^{\lambda}_{\mu\nu}$ denotes
the Christoffel symbols, and $\gamma=\psi^{12}r^4\sin^2\theta$ is the determinant of the 
three-metric, $\gamma_{i j} = \psi^4 f_{i j}$. The conserved rest-mass
density $D$, momentum density in the $j$-direction $S_j$, and the
total energy density $E$ are defined as
\begin{equation}
D=\rho W\ ,\quad S_{j} = \rho h W^2 \nu_j\ ,\quad E=\rho h W^2 - P\ ,
\end{equation}
where $W=1/\sqrt{1-\nu_i\nu^i}$ is the Lorentz factor, $\nu^i$ and $\nu^{*i}$
represent the Eulerian and ``advective" velocities, in the spherically-symmetric 
case equal to $u^i/W$ and $\alpha u^i/W$, respectively.

As in \citet{torres:17}, we consider the linear perturbations of the system
with respect to the equilibrium static background, for which the only
non-zero radial component of Eq.~\ref{hydro2} is
\begin{equation}
\frac{1}{\rho h}\partial_r P = -\partial_r \ln\alpha \equiv G_r\ ,
\end{equation}
where $G_r\equiv \tilde{G}$ is the radial component of the gravitational acceleration.
At the same time, in addition to the perturbation of density, pressure, and velocity, we
introduce the non-zero perturbation of the lapse function, $\alpha$. This addition does not
fully relax the Cowling approximation, but it closely mimics
the conditions of our numerical simulations, where the shift vector $\beta^i=0$ and the conformal
factor is fixed $\psi=1$. Following \citet{torres:17}, we denote the
Eulerian perturbations of the quantities by $\delta$ and the
Lagrangian perturbations by $\Delta$, where the relation between the
two for any quantity, e.g. $\rho$, is 
\begin{equation}
\Delta\rho=\delta\rho + \xi^i\partial_i\rho\ .
\end{equation}
Here, $\xi^i$ is the Lagrangian displacement of a fluid element, related to the
advective velocity as
\begin{equation}
\partial_t \xi^i = \delta \nu^{*i}\ .
\end{equation}
After perturbing the quantities by substituting, e.g., $\rho\rightarrow\rho+\delta\rho$,
and leaving only terms of linear order, Eqs.~\ref{hydro1} and~\ref{hydro2} can be
rewritten as
\begin{eqnarray}
&&\label{hydro11}\frac{\Delta \rho}{\rho} = -\left(\partial_i\xi^i + \xi^i \partial_i \ln\sqrt{\gamma}\right)\ , \\
&&\label{hydro22}\rho h \partial_t \delta\nu_j + \alpha \partial_j \delta P + \delta\alpha \partial_j P = 
- \delta(\rho h) \partial_j \alpha - \rho h \partial_j \delta\alpha\ ,
\end{eqnarray}
with the three components of Eq.~\ref{hydro22} taking the form
\begin{eqnarray}
&&\label{hydro111}\rho h \psi^4\alpha^{-2}\frac{\partial^2\xi^r}{\partial t^2}+\partial_r \delta P +
\frac{\delta\alpha}{\alpha}\partial_r P = \delta(\rho h) \tilde{G} - \frac{\rho h}{\alpha}\partial_r \delta\alpha\ , \\
&&\label{hydro222}\rho h \psi^4\alpha^{-2}r^2\frac{\partial^2\xi^{\theta}}{\partial t^2}+\partial_{\theta} \delta P = 
- \frac{\rho h}{\alpha}\partial_{\theta} \delta\alpha\ , \\
&&\label{hydro333}\rho h \psi^4\alpha^{-2}r^2\sin^2\theta\frac{\partial^2\xi^{\phi}}{\partial t^2}+\partial_{\phi} \delta P = 
- \frac{\rho h}{\alpha}\partial_{\phi} \delta\alpha\ .
\end{eqnarray}

The condition of adiabaticity of the perturbations
\begin{equation}
\label{adiabat}
\frac{\Delta P}{\Delta \rho} = h c_s^2 = \frac{P}{\rho}\Gamma_1\ ,
\end{equation}
where $c_s$ is the relativistic speed of sound and $\Gamma_1$ is the
adiabatic index, allows one to write \citep{torres:17}:
\begin{equation}
\label{deltarhoh}
\delta(\rho h) = \left(1+\frac{1}{c_s^2}\right)\delta P - \rho h \xi^i \mathcal{B}_i\ ,
\end{equation}
where
\begin{equation}
\mathcal{B}_i \equiv \frac{\partial_i e}{\rho h} - \frac{1}{\Gamma_1}\frac{\partial_i P}{P}
\end{equation}
is the relativistic version of the Schwarzschild discriminant
and $e\equiv \rho(1+\epsilon)$. For a spherically symmetric background, the
only non-zero component of $\mathcal{B}_i$ is $\mathcal{B}_r=\mathcal{B}$.
Due to the adiabatic nature of perturbations, Eq.~\ref{hydro3}
does not add any information.

To close the system of Eqs.~\ref{hydro1}-\ref{hydro2}, we use the Poisson equation
\begin{equation}
\nabla^2 \delta \Phi = 4\pi \delta \rho\ ,
\end{equation}
where $\Phi$ is the gravitational potential. Using the relation $\alpha=e^{\Phi}$ we
rewrite it as
\begin{equation}
\label{poisson}
\nabla^2 \left( \frac{\delta\alpha}{\alpha}\right) = 4\pi \delta\rho\ .
\end{equation}

Following \citet{torres:17}, we consider only polar perturbations and 
expand them in terms of spherical harmonics as
\begin{eqnarray}
&&\delta P = \delta \hat{P}\,Y_{lm}e^{-i\sigma t}\ , \quad 
\delta \alpha = \delta \hat{\alpha}\,Y_{lm}e^{-i\sigma t}\ , \nonumber \\
&&\xi^r=\eta_r \,Y_{lm}e^{-i\sigma t}\ , \quad
\xi^{\theta}=\eta_{\bot}\frac{1}{r^2}\,\partial_{\theta}Y_{lm}e^{-i\sigma t}\ ,
\end{eqnarray}
where $\delta \hat{P}$, $\delta \hat{\alpha}$, $\eta_r$, and $\eta_{\bot}$
are scalar functions depending only on radial coordinate. With this
ansatz, and using the adiabaticity condition~\ref{adiabat}, 
Eq.~\ref{poisson} may be brought to the form
\begin{equation}
\frac{1}{r^2}\frac{\partial}{\partial r}r^2\frac{\partial}{\partial r}\left(\frac{\delta\hat{\alpha}}{\alpha}\right)
- \frac{1}{\alpha}\frac{l(l+1)}{r^2}\delta\hat{\alpha} = 
4\pi\left[\frac{\rho}{P\Gamma_1}\left(\delta\hat{P}+\eta_r\partial_r P\right)
-\eta_r \partial_r\rho\right]\ .
\end{equation}
To conveniently find the numerical solution, we introduce the
function $f_{\alpha}=\partial_r(\delta\hat{\alpha}/\alpha)$ and break this second-order equation 
into two first-order equations:
\begin{eqnarray}
&&\label{pois1}\frac{2}{r}f_{\alpha}+\partial_r f_{\alpha} - \frac{1}{\alpha}\frac{l(l+1)}{r^2}\delta\hat{\alpha} = 
4\pi\left[\frac{\rho}{P\Gamma_1}\left(\delta\hat{P}+\eta_r\partial_r P\right)
-\eta_r \partial_r\rho\right]\ , \\
&&\label{pois2}-\frac{\partial_r\alpha}{\alpha^2}\delta\hat{\alpha} +
\frac{1}{\alpha}\partial_r \delta\hat{\alpha} = f_{\alpha}\ .
\end{eqnarray}

Eq.~\ref{hydro222} results in
\begin{equation}
\label{deltaP}
\delta\hat{P} = q \sigma^2\eta_{\bot} - \frac{\rho h}{\alpha}\delta\hat{\alpha}\ ,
\end{equation}
where, after \citet{torres:17}, we have defined $q\equiv \rho h \alpha^{-2}\psi^4$.
Using Eqs.~\ref{adiabat}, \ref{deltarhoh}, and~\ref{deltaP} in Eqs.~\ref{hydro11} 
and~\ref{hydro111}, we get
\begin{eqnarray}
&&\label{fineq1}\partial_r \eta_r + \left[\frac{2}{r}+\frac{1}{\Gamma_1}\frac{\partial_r P}{P} +
6\frac{\partial_r \psi}{\psi}\right]\eta_r + \frac{\psi^4}{\alpha^2 c_s^2}
\left(\sigma^2-\mathcal{L}^2\right)\eta_{\bot}
-\frac{1}{\alpha c_s^2}\delta\hat{\alpha} = 0\ , \\ 
&&\label{fineq2}\partial_r \eta_{\bot} - \left(1-\frac{\mathcal{N}^2}{\sigma^2}\right)\eta_r 
+ \left[\partial_r\ln q-\tilde{G}\left(1+\frac{1}{c_s^2}\right)\right]\eta_{\bot} 
-\frac{1}{\alpha\tilde{G}}\frac{\mathcal{N}^2}{\sigma^2}\delta\hat{\alpha} = 0\ ,
\end{eqnarray}
where $\mathcal{N}$ is the relativistic Brunt-V\"{a}is\"{a}l\"{a} frequency 
defined as 
\begin{equation}
\mathcal{N}^2\equiv \frac{\alpha^2}{\psi^4}G^i \mathcal{B}_i = \frac{\alpha^2}{\psi^4}\tilde{G} \mathcal{B}
\end{equation}
and $\mathcal{L}$ is the relativistic Lamb frequency
\begin{equation}
\mathcal{L}^2\equiv \frac{\alpha^2}{\psi^4}c_s^2\frac{l(l+1)}{r^2}\ .
\end{equation}
Finally, using Eq.~\ref{deltaP}, we bring Eqs.~\ref{pois1} and~\ref{pois2} to the
form
\begin{eqnarray}
&&\label{fineq3}\partial_r f_{\alpha} +\frac{2}{r}f_{\alpha} + 4\pi \left[\partial_r\rho - 
\frac{\rho}{P\Gamma_1}\partial_r P\right]\eta_{r} - \frac{4\pi\rho}{P\Gamma_1}
q\sigma^2\eta_{\bot}+\left[\frac{4\pi\rho^2 h}{P\Gamma_1\alpha}-
\frac{1}{\alpha}\frac{l(l+1)}{r^2}\right]\delta\hat{\alpha} = 0\ , \\
&&\label{fineq4} \partial_r \delta\hat{\alpha} = f_{\alpha}\alpha + 
\frac{\partial_r \alpha}{\alpha}\delta\hat{\alpha}\ .
\end{eqnarray}

To find the eigenfrequencies of the linear perturbation modes, $f=\sigma/(2\pi)$,
we numerically solve the system of first-order differential equations 
\ref{fineq1}, \ref{fineq2}, \ref{fineq3}, and~\ref{fineq4}.

\newpage

\section{Results for $l=3$ and $l=4$ modes}
\label{Ap2}

\begin{figure*}[!htb]
  \centering
  \includegraphics[width=0.95\textwidth]{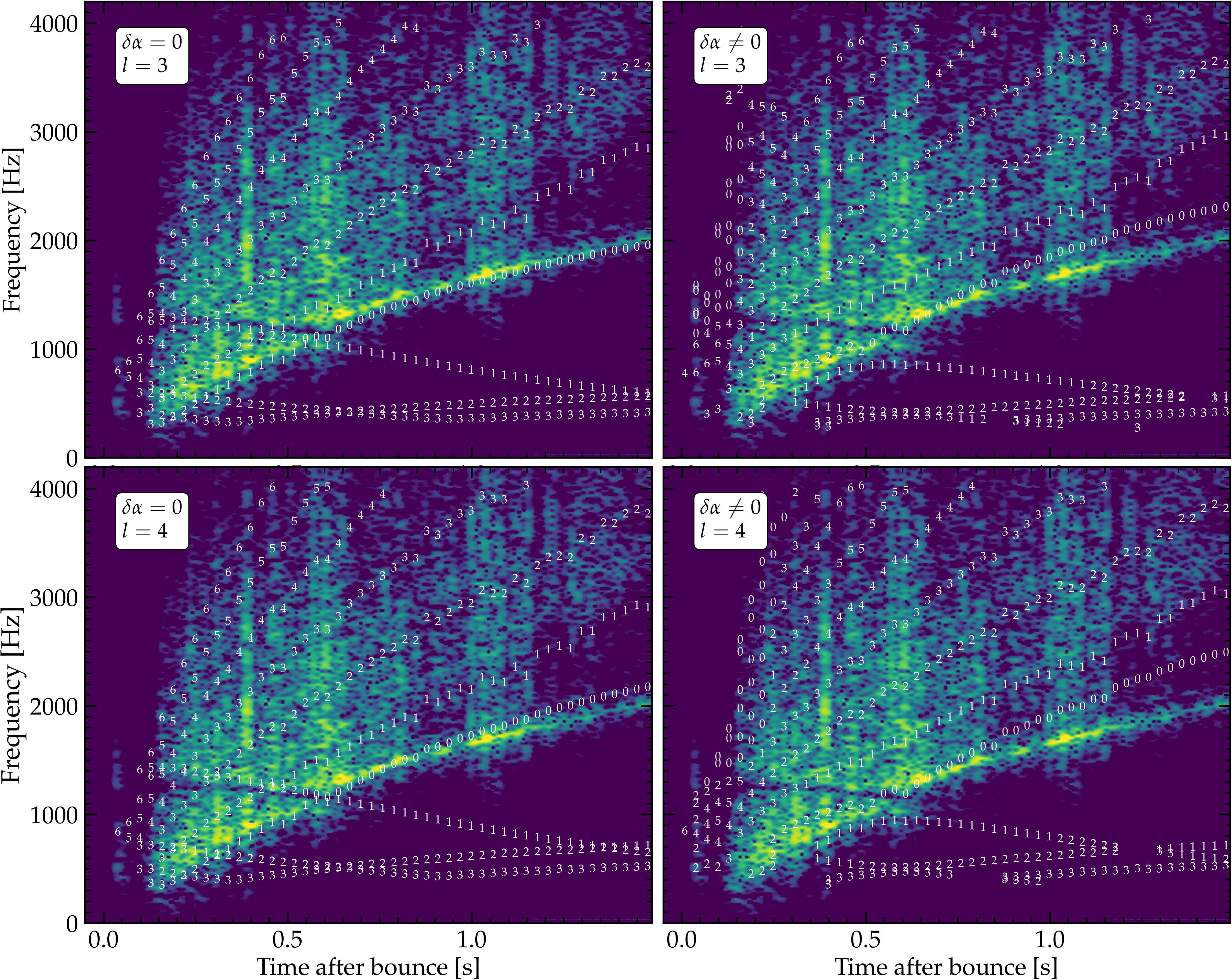}
  \caption{Eigenfrequencies $\sigma/2\pi$ of the $l=3$ (top two panels) and $l=4$ 
  (bottom two panels) modes compared to the GW spectrogram
  for the model \texttt{M10\_SFHo}. Each digit represents the number of nodes in the 
  corresponding mode. The left panels show the results obtained using the
  Cowling approximation, while the right panels show the solution of the full system of 
  Eqs.~(\ref{fineq1t})-(\ref{fineq4t}). The fundamental $l=3$ mode in the top left panel
  seems to coincide with the dominant GW frequency, but it shifts upwards once we relax
  the Cowling approximation in the top right panel.} 
  \label{fig:l3l4}
\end{figure*}

\end{document}